\documentclass{elsart}

\usepackage[square,sort&compress,comma]{natbib}

\usepackage{graphicx}% eps files
\usepackage{bm}% bold math
\usepackage{amsmath}% equation formatting
\usepackage{amsfonts}% equation formatting
\newlength{\twocolumnwidth}
 
\newcommand{\preprintmargin}{ } 
\newcommand{\Eq}[1]{Eq.\ (\protect\ref{eq:#1})}
\newcommand{\Eqs}[1]{Eqs.\ (\protect\ref{eq:#1})}

\newcommand{\hide}[1]{} 
\newcommand{\eqlabel}[1]{\addtocounter{equation}{1}
	\tag{\arabic{equation}{\scriptsize \{eq:#1\}}}\label{eq:#1}} 
\newcommand{\chlabel}[1]{\{#1\}\label{#1}} 
\newcommand{\Heis}[1]{\protect{\hat {\mathcal #1}}} 
 
\newcommand{\R}{_{\text{R}} } 
\newcommand{\FP}{^{(+)}} 
\newcommand{\FM}{^{(-)}}

\newcommand{\av}[1]{\overline{\hspace{0.1ex}#1\hspace{0.1ex}}} 
 
\newcommand{\Tr}{\text{Tr}}

\renewcommand{\bm}[1]{{\mbox{\rm\boldmath$#1$}}}

\newcommand{\generalbracket}[4]{\protect\ensuremath{#1#2#4#1#3}} 
 
\newcommand{\Sbracket}[2]{\protect\generalbracket{#1}{[}{]}{#2}} 
\newcommand{\Cbracket}[2]{\protect\generalbracket{#1}{\{}{\}}{#2}} 
\newcommand{\Qbracket}[2]% 
 {\protect\generalbracket{#1}{\langle}{\rangle}{#2}} 
\newlength{\QQlength}
 
\newcommand{\Vbracket}[2]% 
 {\protect\generalbracket{#1}{\langle 0#1|}{|0#1\rangle}{#2}} 
\newcommand{\vacavbracket}[2]% 
 {#1\langle 0#1|#2#1|0#1\rangle} 
\newcommand{\A}{ }
\newcommand{\formA}[1]%
 {{\renewcommand{\A}{&}\begin{aligned}#1\end{aligned}}}
\newcommand{\formG}[1]% 
 {{\renewcommand{\A}{ }\begin{gathered}#1\end{gathered}}}
\newcommand{\eqM}[2]% 
 {{\renewcommand{\A}{ }\begin{multline}\preprintmargin #1\preprintmargin #2\end{multline}}}
\newcommand{\eqMW}[2]% 
 {{\renewcommand{\A}{ }\begin{multline}#1 #2\end{multline}}}
\newcommand{\eqA}[2]% 
 {\protect{\begin{align}{{\renewcommand{\A}{&}\begin{aligned}#1\end{aligned}}}#2\end{align}}}
\newcommand{\eqG}[2]% 
 {\protect{\begin{gather}{{\renewcommand{\A}{ }\begin{gathered}#1\end{gathered}}}#2\end{gather}}}
 
\newcommand{\qav}[1]{\left\langle #1\right\rangle} 
\newcommand{\nav}[1]{\left\langle \bm{:} #1 \bm{:} \right\rangle} 
\newcommand{\tnav}[1]{\left\langle {\mathcal T} \bm{:} #1 \bm{:} \right\rangle} 
\newcommand{\ket}[1]{\left|#1\right\rangle}

\newcommand{\melc}[3]
	{\left\langle #1\left|#2\right|#3\right\rangle}

\newcommand{\dg}{^{\dag}} 
\newcommand{\dgp}[1]{^{\dag #1}} 
 
\newcommand{\ULM}{Abteilung Quantenphysik, Universit\"at Ulm, 
D-89069 Ulm, Germany} 
\newcommand{\hcQ}{{\hat{{\mathcal Q}}}} 
\renewcommand{\eqlabel}[1]{\label{eq:#1}}
\renewcommand{\chlabel}[1]{\label{#1}}

\newcommand{\StigAffilSw}{Physics Department, Royal Institute of Technology, KTH, Stockholm, Sweden}
\newcommand{\StigAffilFin}{Laboratory of Computational Engineering, HUT, Espoo, Finland}
\newcommand{\U}{\Heis{U}}
\newcommand{\eqOscDFbyDR}{Eq.\ (48) of \cite{RespOsc}}
\newcommand{\eqOscDbyDR}{Eq.\ (47) of \cite{RespOsc}}
\newcommand{\chFieldN}{section 3.1 of \cite{RespOsc}}
\newcommand{\chFieldC}{section 3.2 of \cite{RespOsc}}
\newcommand{\TpmVsTimeLoop}{section 2.2 of Ref.\ \cite{RespOsc}}
\newcommand{\chKuboSchwinger}{section 4.3.1 of Ref.\ \cite{RespOsc}}
\newcommand{\DiscEndSecII}{in section 2 of Ref.\ \cite{RespOsc}}  
\begin{document}

\begin{frontmatter}



\title{Causal signal transmission by quantum fields.\\  
II.\ Quantum-statistical response of interacting bosons}


%*******************************************
\author[a1]{L.\ I.\ Plimak} 
%*******************************************
\author[a1,a2,a3]{and S.\ Stenholm} 
\address[a1]{\ULM.} 
\address[a2]{\StigAffilSw.} 
\address[a3]{\StigAffilFin.} 
%******************************************* 

\begin{abstract}
% Text of abstract
We analyse nonperturbatively 
signal transmission patterns in Green's functions of  
interacting quantum fields.  
Quantum field theory is re-formulated 
in terms of the nonlinear quantum-statistical response of the field. 
This formulation applies equally to interacting relativistic fields and 
nonrelativistic models. 
Of crucial importance is that all causality properties to be expected 
of a response formulation indeed hold. 
Being by construction equivalent to Schwinger's closed-time-loop formalism, this formulation is also shown to be related naturally to both Kubo's 
linear response and Glauber's macroscopic photodetection theories, 
being a unification of the two with generalisation to the 
nonlinear quantum-statistical response problem. 
In this paper we introduce response formulation of bosons; 
response reformulation of fermions will be subject of a separate paper. 
\end{abstract}


\end{frontmatter}

% main text
%******************************************* 
\section{Introduction} 
%******************************************* 
This paper continues the investigation of response properties of quantum systems started in Ref.\ \cite{RespOsc}. 
In \cite{RespOsc} we introduced a {\em response formulation\/} of the harmonic oscillator and extended it to noninteracting bosonic fields. 
Here we show that response formulation may be further extended to arbitrary interacting bosonic fields. 
Response reformulation of fermions will be subject of a separate paper. 

For noninteracting bosons \cite{RespOsc}, response formulation means  description of the quantum system in terms of quantum averages of the {\em normally-ordered products\/} \cite{Schleich,MandelWolf,KellyKleiner,GlauberTN} of field operators defined in the presence of external sources. In \cite{RespOsc}, we proved that this description is equivalent to the standard quantum-field-theoretical description of the same system within Schwinger's renown closed-time-loop formalism \cite{SchwingerC}. The response formulation and the Schwinger formalism are coupled by a one-to-one {\em response substitution\/} in the corresponding characteristic functionals.

It would seem that the obvious way of generalising this result to interacting fields is replacing the normal ordering of free-field operators by the {\em time-normal\/} ordering of Heisenberg operators as introduced by Glauber and Kelly and Kleiner \cite{KellyKleiner,GlauberTN}. However, we quickly discover that this leads to loss of the key property of the response formulation: its equivalence to the standard Green-function approach. Without an amendment to the concept of time-normal ordering, response substitution for interacting bosons does not exist. This makes the whole exercise pointless: recall that our ultimate goal is extending the phase-space techniques to relativistic problems. We are therefore forced to choose a  
different  
approach. We take the response substitution found in \cite{RespOsc} for noninteracting bosons, postulate it for interacting bosons, and consider the consequences. This implies introducing a new definition replacing the familiar time-normal operator ordering. However, we also show that within the optical paradigm (technically, within the approximation of slowly varying amplitudes) this definition coincides with the definition by Glauber and Kelly and Kleiner.
Interpretation of any of the quantum-optical experiments needs not be reconsidered.

Except being a preparatory work for the phase-space approach to relativistic quantum fields, some results of this paper appear to have significance of their own. First and foremost, we demonstrate that all quantum properties of interacting systems may be interpreted in terms of response and self-radiation. As was explained in paper \cite{RespOsc}, one interpretation of our results is proving the equivalence between Schwinger's closed-time-loop formalism \cite{SchwingerC} on the one hand, and a certain generalisation of Kubo's and Glauber's approaches combined on the other. For more details we refer the reader to the introduction of Ref.\ \cite{RespOsc}. All points made there apply not just to the harmonic oscillator, 
but also to any interacting bosonic quantum system.

Perhaps the most interesting result of this paper is a fundamental link between response and noncomutivity of operators. Indication of this connection may be seen already in Kubo's famous formula for the linear response function \cite{Kubo}, where the latter is expressed by the average of the two-time commutator. This feature is shown to hold for the full nonlinear quantum-statistical response of interacting systems. The assumption that operators commute cancels the dependence of the system properties on external sources turning the system into a ``pre-assigned quantum source.''

Another interesting result, which plays only a technical role in our analyses but seems to be important on its own,  is that all response properties of a system are contained in the field operator defined without the external source. In other words, the information contained in the field operator 
(and in the intial condition)
suffices to describe any {\em scattering experiment\/} performed on 
the system, a fact which has never been fully appreciated and which 
has profound consequences for the quantum measurement theory. Again, a known example of this is Kubo's formula where the linear response to a source is specified in terms of the field operators defined without the source.

One technical point merits special mentioning here. All usual problems of the quantum field theory like adiabatically switching the interaction on and off, renormalisations and the like are securely locked away in the assumption that Heisenberg field operators and Green's functions ``may be defined.'' In particular, we do not make any effort at specifying Green's functions at coinciding time arguments because in all practical calculations such specifications emerge as a side-effect of a {\em regularisation procedure\/}. This is the case in the relativistic quantum field theory \cite{Bogol} as well as for simple nonrelativistic models \cite{BWO}. Note that, in the latter case, regularisations may be applied directly to the phase-space equations in the form of specifying a stochastic calculus \cite{BWO}. A similar regularisation procedure will be introduced in our forthcoming papers for relativistic phase-space models.

The paper is organised as follows. In section \ref{ch:Summ}, we introduce the necessary quantum-field-theoretical concepts, such as time orderings of operators, Green's functions and their characteristic functionals. In section \ref{ch:HarmRevi}, we reiterate the key results of paper \cite{RespOsc}. Using these results as leading considerations, we then proceed to defining the response formulation of interacting bosons. In section \ref{ch:NPert} we prove that the two ways of describing the system, in terms of the field operator defined with and without the external source, are formally equivalent. In section \ref{ch:PhysPr}, we define the concepts inherent in the response formulation, such as time-normal operator ordering, quantum-statistical response functions and related characteristic functionals. In particular, we show why the Glauber-Kelly-Kleiner definition of the time-normal ordering must be amended, and why our amendmend is inessential for quantum optics. In section \ref{ch:NOscResp}, we derive formulae relating Green's functions to response functions and {\em vice versa\/}. Among other things, we show that our expression for the linear response function coincides with Kubo's formula, and give explicit demonstration of the link between operator noncommutativity and response. In section \ref{ch:Caus}, we prove explicit causality in the response formulation. In appendix \ref{app:FPN}, we summarise the necessary facts regarding separating the frequency-positive and negative parts of a function. Finally, in appendix \ref{app:ChBos}, we summarize the results from the response formulation of charged bosons, while in the main body of the paper we only treat neutral bosons. 
%******************************************* 
%\newpage
\section{Statement of the problem}\chlabel{ch:Summ}
%******************************************* 
%\subsection{Definitions}\chlabel{ch:Def}
%******************************************* 
\subsection{Neutral bosons}\chlabel{ch:NBos}
%******************************************* 
\subsubsection{Quantum dynamics in the presence of a source}
\begin{figure}[t]
\begin{center}
\includegraphics[width=0.6\columnwidth]{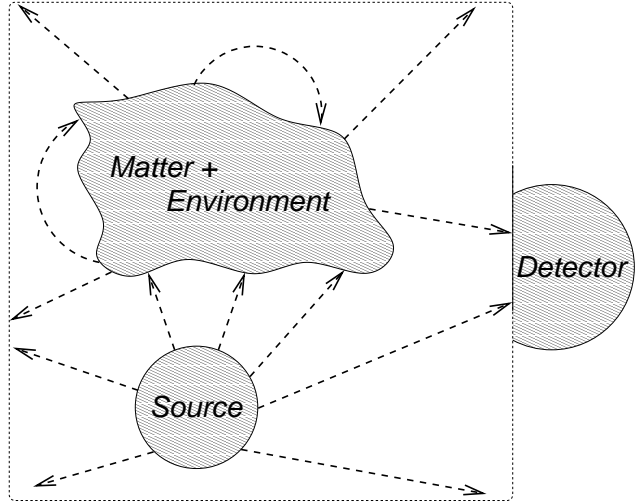}
\end{center}
\caption{% 
Schematic diagram of a response experiment. 
The source and the detector are coupled to a quantum field 
(dashed lines with arrows), e.g., 
the electromagnetic field. 
This field also interacts with some matter and/or environment 
which shapes its observable properties. 
The dotted rectangle encircles the part of the system (field+matter+source) we treat explicitly while a detector is implicit in our 
considerations.%  
}
\label{fig:RespExp}
\end{figure}

%******************************************* 
The Gedanken experiment we have in mind is 
illustrated in Fig.\ \ref{fig:RespExp}. 
It involves two classical devices, the {\em source\/} and the 
{\em detector\/}, coupled to a quantum field described by a 
Hermitian 4-vector 
field operator ${\Heis{Q}}_{\mu }(\bm{r},t)$. 
Formally, we analyse a  
{\em quantum-statistical response of the quantum field\/} 
(cf.\ Fig.\ \ref{fig:RespExp})
by considering a system with  
the generic Hamiltonian  
%=============================================
\eqA{ 
\Heis{H}(t) = \Heis{H}_0(t) + 
\sum_\mu \int d^3\bm{r} j^{\mu}(\bm{r},t){\Heis{Q}}_{\mu}(\bm{r},t) 
\equiv \Heis{H}_0(t) + 
j(t){\Heis{Q}}(t)
.   
}{
\eqlabel{HSource} 
% \nonumber % \eqlabel{} 
}% 
%+++++++++++++++++++++++++++++++++++++++++++++
Here, the external source $j(t)$ is a given real c-number current, and 
we have introduced a general shorthand notation, 
\begin{gather} 
\begin{aligned}
x(t) Y(t) = \sum_{\mu }\int d^3 \bm{r} x^{\mu }(\bm{r},t) Y_{\mu }(\bm{r},t) 
= \sum_{\mu }\int d^3 \bm{r} x_{\mu }(\bm{r},t) Y^{\mu }(\bm{r},t),   
\end{aligned}%
% \nonumber % 
\eqlabel{xY} 
\end{gather}%
where the integration extends the whole 3-dimensional space, cf.\ endnote \cite{IntRange}. 
Shorthand (\ref{eq:xY}) applies by default 
throughout the paper. 

By definition, ${\Heis{Q}}_{\mu}(\bm{r},t)$ and $\Heis{H}_{0}(t)$ are the field operator and the Hamiltonian in the Heisenberg picture with $j=0$:  
\begin{gather} 
\begin{aligned}
i \hbar {\dot {\Heis{Q}}}_{\mu}(\bm{r},t) 
= \big[{\Heis{Q}}_{\mu}(\bm{r},t),\Heis{H}_{0}(t)\big] 
. 
\end{aligned}%
% \nonumber % 
\eqlabel{EqQ} 
\end{gather}%
The field operator and the full Hamiltonian in the Heisenberg picture with $j\neq 0$ will be denoted as ${\Heis{Q}}_{j\mu}(\bm{r},t)$ and $\Heis{H}_j(t)$: 
\begin{align} 
\begin{aligned}
i \hbar {\dot {\Heis{Q}}}_{j\mu}(\bm{r},t) 
= \big[{\Heis{Q}}_{j\mu}(\bm{r},t),\Heis{H}_{j}(t)\big].  
\end{aligned}%
% \nonumber % 
\eqlabel{EqQj} 
\end{align}%
To assign formal meaning to the quanties used in \Eqs{EqQ} and (\ref{eq:EqQj}), we assume the existence of the Schr\"odinger-picture Hamiltonian,   
\begin{gather} 
\begin{aligned}
\Heis{H}_{\text{S}}(t) = 
\Heis{H}_{0\text{S}}(t) + j(t)\Heis{Q}_{\text{S}} 
,  
\end{aligned}%
% \nonumber % 
\eqlabel{HS} 
\end{gather}%
where $\Heis{H}_{0\text{S}}(t) $ includes the free-field Hamiltonian as well as all Hamiltonians describing 
the matter and environment in \mbox{Fig.\ \ref{fig:RespExp}} 
with corresponding  
interactions. 
The interaction of the field with the external source is described by the term $j(t)\Heis{Q}_{\text{S}}$, 
%=============================================
\eqA{ 
j(t)\Heis{Q}_{\text{S}} = \sum_{\mu}\int d^3\bm{r}j^{\mu}(\bm{r},t)\Heis{Q}_{\text{S}\mu}(\bm{r}),
}{
% \nonumber % \eqlabel{} 
}% 
%+++++++++++++++++++++++++++++++++++++++++++++
where $\Heis{Q}_{\text{S}\mu }(\bm{r})$ is the field operator in the Schr\"odinger picture (assumed to be time-independent). 
Then 
\eqA{
\Heis{Q}_{\mu}(\bm{r},t)\A ={\Heis{U}}\dg_0(t)\Heis{Q}_{\text{S}\mu}(\bm{r}){\Heis{U}}_0(t), 
\\
\Heis{Q}_{j\mu}(\bm{r},t)\A ={\Heis{U}}\dg(t)\Heis{Q}_{\text{S}\mu}(\bm{r}){\Heis{U}}(t),   
\\ 
\Heis{H}_{0}(t)\A ={\Heis{U}}\dg_0(t)\Heis{H}_{0\text{S}}(t){\Heis{U}}_0(t), 
\\  
\Heis{H}_{j}(t)\A ={\Heis{U}}\dg(t)\Heis{H}_{\text{S}}(t){\Heis{U}}(t) .   
}{\eqlabel{UjEqs}}%
where the evolution operators $\U(t)$ and $\U_0(t)$ related to $\Heis{H}_{\text{S}}(t)$ and $\Heis{H}_{0\text{S}}(t)$, respectively, obey the Schr\"odinger equations,   
%=============================================
\eqA{ 
i\hbar {\dot{\Heis{U}}}(t) \A = \Heis{H}_{\text{S}}(t) \U(t),  \A 
i\hbar {\dot{\Heis{U}}}_0(t) \A = \Heis{H}_{0\text{S}}(t) \U_0(t). 
}{
% \nonumber % 
\eqlabel{UDef} 
}% 
%+++++++++++++++++++++++++++++++++++++++++++++
It is assumed that $\U(t)\to 1$ as $t\to -\infty$; this applies to all evolution operators defined here as well as in section \ref{ch:NPert}. 

The fact that all nonlinear interactions enter through $\Heis{H}_{0\text{S}}(t) $ in no way prevents us from treating this term as the unperturbed Hamiltonian and the linear term $j(t)\Heis{Q}_{\text{S}}$ as interaction. 
Following the standard time dependent perturbation techniques, we define the interaction-picture evolution operator $\U_j(t)$ as the ``difference'' between $\U(t)$ and $\U_0(t)$, 
\begin{gather} 
\begin{aligned}
\U(t) = \U_0(t)\U_j(t).  
\end{aligned}%
% \nonumber % 
\eqlabel{UUj} 
\end{gather}%
The evolution operator $\U_j(t)$ obeys the equation (using shorthand (\ref{eq:xY}))
%=============================================
\eqA{ 
i\hbar {\dot {\U}}_j(t) &= j(t)\Heis{Q}(t) \U_j(t),  
}{
% \nonumber % \eqlabel{} 
}% 
%+++++++++++++++++++++++++++++++++++++++++++++
solved by the T-exponent, 
\begin{gather} 
\begin{aligned}
{\Heis{U}}_{j}(t) = T_+\exp \left[
-\frac{i}{\hbar } \int_{-\infty}^t dt' j(t')\Heis{Q}(t')
\right] . 
\end{aligned}%
% \nonumber % 
\eqlabel{EqUj} 
\end{gather}%
We can then relate $\Heis{Q}_{j\mu}(\bm{r},t)$ to $\Heis{Q}_{\mu}(\bm{r},t)$ directly, 
%=============================================
\eqA{ 
\Heis{Q}_{j\mu}(\bm{r},t)&={\Heis{U}}_j\dg(t)\Heis{Q}_{\mu}(\bm{r},t){\Heis{U}}_j(t).   
}{
% \nonumber % 
\eqlabel{QjByQ} 
}% 
%+++++++++++++++++++++++++++++++++++++++++++++

For purposes of this paper, it suffices to assume that the Heisenberg operators ${\Heis{Q}}_{\mu}(\bm{r},t)$ ``in the absence of the source'' and ${\Heis{Q}}_{j\mu}(\bm{r},t)$ ``in the presence of the source'' are known. Equations (\ref{eq:EqUj}) and (\ref{eq:QjByQ}) show that ${\Heis{Q}}_{j\mu}(\bm{r},t)$ may be expressed by ${\Heis{Q}}_{\mu}(\bm{r},t)$, so that the only assumption we really need is that the operator ${\Heis{Q}}_{\mu}(\bm{r},t)$ is known. 
All other details, including the actual physical nature of the operator 
${\Heis{Q}}_{\mu}(\bm{r},t)$, are irrelevant. 
Our choice of a 4-vector field 
suggests quantum electrodynamics,  
however, we do not impose gauge conditions nor symmetry nor any transformation 
properties. 
What we do here 
equally applies to a full relativistic bosonic field 
and a single resonator mode. 
The index $\mu $ can in fact be anything, and run over any arbitrary 
set of indices, modes or polarisations. 

Furthermore, we do not make any assumptions regarding the Hamiltonian $\Heis{H}_{0\text{S}}(t)$ except its existence. It may include arbitrary interactions with other quantum fields (cf.\ ``Matter+Environment'' in Fig.\ \ref{fig:RespExp}). These interactions may be explicit, or accounted for phenomenologically as nonlinearities, or treated as heat-baths, or as any combination of these. 
Nothing precludes them from being explicitly time-dependent. 
In other words, our analyses apply to arbitrary open nonlinear 
systems. 
We do not assume that  
{\em closed\/} evolution 
equations of any kind exist for the field operators.  
This frees us of all 
physical restrictions, be it energy conservation or Markovian nature 
of the evolution. 
The only applicable condition is conservation of probability.

\newcommand{\Q}[1]{{\Heis{Q}}_{\mu #1}(\bm{r} #1,t #1)} 
\newcommand{\Qj}[1]{{\Heis{Q}}_{j\mu #1}(\bm{r} #1,t #1)} 
\newcommand{\etp}[2]{\eta_+^{#1\mu #2}(\bm{r} #2,t #2)} 
\newcommand{\etm}[2]{\eta_-^{#1\mu #2}(\bm{r} #2,t #2)} 
\newcommand{\ett}[2]{\eta^{#1\mu #2}(\bm{r} #2,t #2)} 
\newcommand{\srt}[1]{\sum_{\mu #1}\int d^3\bm{r} #1 dt #1} 
%******************************************* 
\subsubsection{Closed-time-loop formalism}
%******************************************* 
The minimal approach capable of accommodating this kind of generality is Schwinger's famous closed-time-loop formalism \cite{SchwingerC}. Our quantities of interest are the Schwinger-Perel-Keldysh-style \cite{Keldysh} double-time-ordered averages of the field operators 
%=============================================
\eqA{ 
\Big\langle T_-\Q{_1}\cdots\Q{_m}\,%\\ \times 
T_+\Q{'_1}\cdots\Q{'_n}\Big\rangle . 
}{
\eqlabel{FGNL} % 
}% 
%+++++++++++++++++++++++++++++++++++++++++++++
Time-ordering $T_+$ ($T_-$) puts operators in order of 
increasing (decreasing) time arguments. By definition, bosonic operators under the ordering sign commute, and, if $t_1<t_2<\cdots<t_m$, 
%=============================================
\eqA{ 
T_-\Heis{Q}(t_1)\cdots\Heis{Q}(t_m)  = 
\Heis{Q}(t_1)\cdots\Heis{Q}(t_{m-1})\Heis{Q}(t_m), 
\\
T_+\Heis{Q}(t_1)\cdots\Heis{Q}(t_m) = 
\Heis{Q}(t_m)\Heis{Q}(t_{m-1})\cdots\Heis{Q}(t_1) 
,   
}{
\eqlabel{TpmDef} 
}% 
%+++++++++++++++++++++++++++++++++++++++++++++
where all other arguments of the field operator are omitted for brevity. 
Equivalence of the double-time ordering with the closed-time-loop ordering was discussed in \TpmVsTimeLoop. 
The quantum averaging in 
(\ref{eq:FGNL}) is defined in the standard way 
with respect to the Heisenberg density matrix $\rho _0$, 
\begin{gather} 
\begin{aligned}
\qav{\cdots} = \Tr \rho_0 (\cdots). 
\end{aligned}%
% \nonumber % 
\eqlabel{QAv0} 
\end{gather}%
This definitions implies the Heisenberg picture with respect to Hamiltonian $\Heis{H}_0$. 
For reasons explained in the introduction, we do not make any effort at specifying Green's functions at coinciding time arguments.  

The whole assemblage of Green's functions (\ref{eq:FGNL}) is conveniently handled through their generating, or characteristic,  
functional $\Phi (\eta _-,\eta _+)$: 
%\begin{widetext} 
\eqMW{
\qav{T_-\Q{_1}\cdots\Q{_m}\,
T_+\Q{'_1}\cdots\Q{'_n}} \\ =  
\frac{(-i)^m i^n \, 
\delta^{m+n}\Phi (\eta _-,\eta _+)}{
\delta\etm{}{_1} 
\cdots 
\delta\etm{}{_m} 
\delta\etp{}{'_1} 
\cdots 
\delta\etp{}{'_n} 
}\Big|_{\eta _-=\eta _+ =0}
\, 
,  
}{\eqlabel{PhiDefDer}}
\eqMW{
\Phi (\eta _-,\eta _+) =  
\Big\langle
T_- \exp \Big [ i \int dt\,\eta _-(t) {\hat{{\mathcal Q}}}(t)\Big ] 
\ T_+ \exp \Big [-i \int dt\,\eta _+(t) {\hat{{\mathcal Q}}}(t)\Big ]
\Big\rangle \\ =  1 + 
\sum_{m+n\geq 1} \frac{i^m (-i)^n }{m!n!} %\\ \times  
\left(\int dt\right)^{m+n} %\\ \times 
\eta _-(t_1)\cdots\eta _-(t_m)
\eta _+(t'_1)\cdots\eta _+(t'_n)\, \\ \times  
\qav{T_-\Heis{Q}(t_1)\cdots\Heis{Q}(t_m)
T_+\Heis{Q}(t'_1)\cdots\Heis{Q}(t'_n)}
. 
}{\eqlabel{PhiDef}}%
%\end{widetext}%
The functional arguments   
$\eta _{\pm}^{\mu}(\bm{r},t)$ here 
are arbitrary smooth 4-vector c-number functions, and the symbol $\left(
\int dt
\right)^{m+n} $ denotes integration over all time arguments. 
Omitting integration limits means that the time integrations are from minus to plus infinity, cf.\ also endnote \cite{IntRange}. 
Equation (\ref{eq:PhiDef}) implies shorthand (\ref{eq:xY}); 
it also gives an example of a ``loose'' usage of this shorthand, 
with the 
corresponding pairs (e.g., $\eta _-(t_1)$ and $\Heis{Q}(t_1)$) separated by other factors.
Definitions identical to (\ref{eq:FGNL})--(\ref{eq:PhiDef}) may also be written with operator $\Qj{}$ in place of $\Q{}$. 
This introduces another functional 
\begin{multline} % 
\preprintmargin
\Phi (\eta _-,\eta _+;j) \\ =  
\Big\langle
T_- \exp \Big [ i \int dt\,\eta _-(t) {\hat{{\mathcal Q}}}_j(t)\Big ] 
\,%\\ \times  
T_+ \exp \Big [-i \int dt\,\eta _+(t) {\hat{{\mathcal Q}}}_j(t)\Big ]
\Big\rangle ,  
\eqlabel{PhiDefJ} 
\preprintmargin
\end{multline}% 
so that $ 
\Phi (\eta _-,\eta _+) =\Phi (\eta _-,\eta _+;0)  
$. 
For simplicity, we assume that 
\begin{gather} 
\begin{aligned}
j(t) = 0, \ \ |t|>t_{\text{max}} ,  
\end{aligned}%
% \nonumber % \eqlabel{} 
\end{gather}%
where $t_{\text{max}}$ is some large parameter, and that the Heisenberg pictures with and without the external source coincide for $t<-t_{\text{max}} $. This allows us to use the same averaging in (\ref{eq:PhiDef}) and (\ref{eq:PhiDefJ}). 

Shorthand (\ref{eq:xY}) plays in our analyses much more important a role than merely saving space. It effectively reduces the problem of the quantum field to the generic case of nonlinear quantum oscillator interacting with environment, with $\Heis{Q}(t)$ being the 
oscillator coordinate. 
In turn, the 
analyses introduced with the nonlinear oscillator suffice because our formal manipulations 
will always apply to the time argument of the field operator leaving the rest of the field arguments (labels) alone. 
All formulae in the analyses below belong to one of two groups. 
Firstly, {\em relations for characteristic functionals\/} like (\ref{eq:PhiDef}) apply in the field case directly by virtue of definition (\ref{eq:xY}). 
Secondly, 
{\em relations for operators and quantum averages\/} like (\ref{eq:TpmDef}) are generalised by simply restoring all labels
\begin{gather} 
\begin{aligned}
t \to \mu , \bm{r},t, \ \ \int dt \to \sum_{\mu }\int d^3\bm{r} dt . 
\end{aligned}%
% \nonumber % 
\eqlabel{ToF} 
\end{gather}%
We shall consistently refer to $\Heis{Q}(t)$ as ``field operator'' so 
as to emphasise that our results apply not just to the nonlinear oscillator but to an arbitrary bosonic field. 
However, treating time as {\em variable\/} and space as {\em label\/} 
makes our viewpoint essentially nonrelativistic. 
All formulae in this paper apply, strictly speaking, in a given 
reference frame. 
%******************************************* 
\subsection{Charged fields}\chlabel{ch:ChBos}
%******************************************* 
A \textit{charged field} is described by introducing a Hermitian-conjugated
pair $ \mathcal{F}_{\mu }(\mathbf{r},t),\mathcal{F}_{\mu }^{\dagger }(%
\mathbf{r},t)$ of Heisenberg operators where%
\begin{align} 
\begin{aligned}
\Heis{F}\dg_{\mu }(\bm{r},t) = \big[\Heis{F}_{\mu }(\bm{r},t)\big]\dg
. 
\end{aligned}%
% \nonumber % \eqlabel{} 
\end{align}% 
Above definitions for neutral 
bosons can be 
generalised to a charged bosonic field by simply 
redefining the linear forms of field operators: 
\begin{align} 
j(t){\Heis{Q}}(t) &\equiv  %\\ 
\eqlabel{HintRedef} 
\sum_\mu \int d^3\bm{r}\Big[j^{\mu\ast}(\bm{r},t)
{\Heis{F}}_{\mu}(\bm{r},t) + 
{\Heis{F}}\dg_{\mu}(\bm{r},t) j^{\mu}(\bm{r},t)\Big],  
\\ 
\eta_{\pm}(t){\Heis{Q}}(t) &\equiv %\\ 
\sum_\mu \int d^3\bm{r}\Big[ 
\bar \eta_{\pm} ^{\mu}(\bm{r},t){\Heis{F}}_{\mu}(\bm{r},t) +  
{\Heis{F}}\dg_{\mu}(\bm{r},t)\eta_{\pm} ^{\mu}(\bm{r},t)\Big]  
,    
\eqlabel{EtaQRedef} 
\end{align}%
cf.\ endnote \cite{IntRange}. This equals treating   
$\Heis{F},\Heis{F}\dg$ as a two-component field, 
or, putting it another way, introducing an additional ``label  
of complexity'' distinguishing $\Heis{F}$ from $\Heis{F}\dg$. 
Replacing $\Heis{Q}$ by a pair of operators 
leads to doubling all functional arguments. 
Functional (\ref{eq:PhiDef}) is replaced by 
%%\begin{widetext} 
\begin{multline} % 
\preprintmargin
\Phi (\eta _-,\bar \eta _-,\eta _+,\bar \eta _+) =  %\\ 
\Qbracket{\Big}{ 
T_- \exp\left\{
i \int dt  \Big[ 
\bar \eta_- (t){\Heis{F}}(t)+{\Heis{F}}\dg(t)\eta_-(t) 
\Big] 
\right\} \,
\\ \times  T_+  \exp\left\{
- i \int dt  \Big[ 
\bar \eta_+ (t){\Heis{F}}(t)+{\Heis{F}}\dg(t)\eta_+(t) 
\Big] 
\right\} 
} 
.      
% 
% \nonumber % 
\eqlabel{DefPhiC} % 
\preprintmargin
\end{multline}% 
%%\end{widetext} 
The definition of Green's functions (\ref{eq:FGNL}) is amended 
accordingly. 
Given the triviality of generalisation to charged bosons, 
and to prevent our formulae from becoming even bulkier, 
in the main body of the paper we only consider neutral bosons. 
Summary of the formulae for charged bosons is given in appendix \ref{app:ChBos}.

The case of fermions (which are always charged) may be reduced 
to the case of charged bosons by redefining operator forms 
(\ref{eq:HintRedef}) and (\ref{eq:EtaQRedef}) with the $\eta $s 
and $j$s being {\em g-numbers\/} in 
place of c-numbers. 
G-numbers, or {\em Grassmann variables\/}, are often 
called {\em classical anticommuting quantities\/}, meaning that 
g-numbers mutually anticommute as c-numbers mutually commute. 
However, introducing Grassmann 
currents into Hamiltonians 
requires a number of modifications of the conventional Hilbert space 
techniques. 
Another subtlety is that definitions of 
Green's functions for fermionic fields 
involve sign conventions. 
Response properties of fermions will therefore be a subject of a separate paper.
%******************************************* 
\section{Harmonic oscillator revisited}\chlabel{ch:HarmRevi}
%******************************************* 
Here we briefly summarise the response formulation of the harmonic oscillator introduced in Ref.\ \cite{RespOsc}. For the oscillator, $\hcQ(t)\to \hat q(t)$ and $\hcQ_j(t)\to \hat q_j(t)$, where $\hat q(t)$ and $\hat q_j(t)$ are the Heisenberg-picture displacement operators defined with and without the source. 
Definitions (\ref{eq:PhiDef}) and (\ref{eq:PhiDefJ}) then apply with the replacements $\hcQ(t)\to \hat q(t)$ and $\hcQ_j(t)\to \hat q_j(t)$, and do not assume shorthand (\ref{eq:xY}). We do not introduce any special notation for the functionals $\Phi (\eta _-,\eta _+)$ and $\Phi (\eta _-,\eta _+;j)$ for the harmonic oscillator. Instead, the fact that a particular relation applies only in the oscillator case will always be stated explicitly. In this section, 
all formulae apply to the harmonic oscillator.  

The response formulation of the oscillator is based on the functional $\Phi\R (\eta ;j)$ defined as  
\begin{gather} 
\begin{aligned}
\Phi \R(\eta ;j) = \nav{
\exp\int dt\, \eta (t)
\hat q_j (t) 
}. 
\end{aligned}%
% \nonumber % 
\eqlabel{qjnorm} 
\end{gather}%
Operators  $\hat q(t)$ and $\hat q_j(t)$ differ by a c-number displacement under the influence of the external source, 
\begin{gather} 
\begin{aligned}
\hat q_j(t) = \hat q(t) + q_j(t), \ \ 
q_j(t) =  \int dt' D\R(t-t')j(t') ,  
\end{aligned}%
% \nonumber % 
\eqlabel{Oscqj} 
\end{gather}%
where $D\R(t)$ is, up to the overall sign, the {\em retarded Green's function\/} of the classical oscillator. The normal ordering is defined primarily for $\hat q(t)$, but, since the difference between it and $\hat q_j(t)$ is a c-number, it applies equally to $\hat q_j(t)$. By making use of (\ref{eq:Oscqj}), \Eq{qjnorm} may be written as 
\begin{gather} 
\begin{aligned}
\Phi \R(\eta ;j) = \Phi_{\text{cl}} (\eta ;j)\Phi_{\text{in}} (\eta),  
\end{aligned}%
% \nonumber % \eqlabel{} 
\end{gather}%
where 
%=============================================
\eqA{ 
\Phi_{\text{cl}} (\eta ;j) = \exp\int dt \,\eta (t)q_j(t) = \exp \int dtdt'\eta (t)D\R(t-t')j(t'),  
}{
% \nonumber % \eqlabel{} 
}% 
%+++++++++++++++++++++++++++++++++++++++++++++
and 
\begin{gather} 
\begin{aligned}
\Phi_{\text{in}} (\eta) = \nav{\exp\int dt \,\eta (t)\hat q(t)}. 
\end{aligned}%
% \nonumber % \eqlabel{} 
\end{gather}%
This way, in the response formulation, response properties of the oscillator separate from those of the initial state. Moreover, only the latter may be quantum. The functional $\Phi_{\text{cl}} (\eta ;j)$ describing the quantum response of the oscillator is by itself a fully classical object. It is a characteristic functional for products of a classical c-number field emitted by a classical c-number current in accordance with laws of classical mechanics: 
\begin{gather} 
\begin{aligned}
q_j(t_1)\cdots q_j(t_m) = 
\frac{\delta^m \Phi_{\text{cl}} (\eta ;j)}
{\delta\eta(t_1)\cdots   \delta\eta(t_m)  } \Big | _{\eta =0}. 
\end{aligned}%
% \nonumber % \eqlabel{} 
\end{gather}%
In the response formulation, the quantum oscillator allows the most classical interpretation possible, assuming that this interpretation remains {\em fully and consistently\/} within the laws of quantum mechanics.

The key result of paper \cite{RespOsc} is that, for the harmonic oscillator, the response formulation is formally equivalent with Schwinger's closed-time-loop Green's function formulation. In \cite{RespOsc}, the following fundamental relation between the characteristic functionals was derived, 
\begin{gather} 
\begin{aligned}
\Phi (\eta _-,\eta _+;j) = \Phi \R(\eta ,\sigma +j),  
\end{aligned}%
% \nonumber % 
\eqlabel{PhiJbyPhiR} 
\end{gather}%
where 
the two pairs of functional variables, $\eta _{\pm}$ and $\eta ,\sigma $, are coupled by an invertible transformation, 
\begin{gather} 
\begin{aligned}
\eta (t) & = -i\big[\eta _+(t)-\eta _-(t)\big], \\  
\sigma(t) & = \hbar \left[
\eta^{(+)} _+(t) + \eta^{(-)} _-(t)
\right],  
\end{aligned}%
% \nonumber % 
\eqlabel{OscSubstEta} 
\\ 
\begin{aligned} 
\eta_+(t) & = i\eta^{(-)} (t) + \frac{1}{\hbar} \sigma(t), \\    
\eta_-(t) & = - i\eta^{(+)} (t) + \frac{1}{\hbar} \sigma(t) ,   
\end{aligned}% 
\eqlabel{OscSubstSigma} 
% \nonumber % \eqlabel{} 
\end{gather}%
Here, 
$g^{(\pm)}(t)$ denote the frequency-positive 
and negative parts 
of a function $g(t)$, defined by dropping the corresponding half 
(negative or positive) of its Fourier-spectrum 
\begin{align} 
\begin{aligned}
g^{(\pm)}(t) = \int \frac{d\omega }{2 \pi } \,
\e{-i\omega t}\theta(\pm \omega )g_{\omega } , 
\ \   
g_{\omega } = \int dt\, \e{i\omega t} g(t).   
\end{aligned}%
% \nonumber % 
\eqlabel{FPNDef} 
\end{align}%
We use the terminology of quantum optics;  
in the quantum field 
theory, the terms frequency-positive and frequency-negative 
are often swapped \cite{Bogol}.  
The concept of the frequency-positive and negative parts 
will be used throughout this paper; 
the necessary facts regarding it are summarised in appendix \ref{app:FPN}. 
Getting back to the characteristic functionals, with $j=0$ equation (\ref{eq:PhiJbyPhiR}) becomes, 
\begin{gather} 
\begin{aligned}
\Phi (\eta _-,\eta _+) = \Phi \R(\eta ,\sigma).   
\end{aligned}%
% \nonumber % 
\eqlabel{PhibyPhiR} 
\end{gather}%

For purposes of this paper, the characteristic property of the response formulation of the quantum oscillator is that it is {\em structurally identical\/} with the response picture of a classical oscillator. For the latter, the general solution for the displacement is 
\begin{gather} 
\begin{aligned}
q(t) = q_{\text{in}}(t) + q_j(t),  
\end{aligned}%
% \nonumber % \eqlabel{} 
\end{gather}%
where $q_{\text{in}}(t)$ is the free oscillation for $j=0$, and $q_j(t)$ is defined by (\ref{eq:Oscqj}). If the initial condition is random, the classical oscillator is characterised by the stochastic moments of the displacement, $\av{q(t_1)\cdots q(t_m)}$, with the bar denoting the classical stochastic averaging. The characteristic functional for these is 
\begin{gather} 
\begin{aligned}
\Phi \R(\eta ;j) = \av{\exp\int dt\,\eta (t)\left[
q_{\text{in}}(t) + q_j(t)
\right] }. 
\end{aligned}%
% \nonumber % 
\eqlabel{PhiRcl} 
\end{gather}%
Denoting the classical functional (\ref{eq:PhiRcl}) by $\Phi \R(\eta ;j)$ is perfectly justified, because \Eq{PhiRcl} holds equally for the quantum functional (\ref{eq:qjnorm}). For the latter, $q_{\text{in}}(t)$ must be defined as a matrix element over a coherent state $\ket{\alpha}$, 
\begin{gather} 
\begin{aligned}
q_{\text{in}}(t) = \melc{\alpha}{\hat q(t)}{\alpha},  
\end{aligned}%
% \nonumber % \eqlabel{} 
\end{gather}%
and the bar should be interpreted as an averaging over the P-function corresponding to the initial state. If the P-function is positive, the classical and quantum response pictures are indistinguishable; with a nonpositive P-function they remain structurally identical. 
%******************************************* 
\section{The Kubo and Schwinger currents for interacting bosons}\chlabel{ch:NPert}
%******************************************* 
Our first step towards the generalisation of the results for the harmonic oscillator to interacting bosons will be finding a replacement for \Eq{Oscqj} expressing the operator ``with the source'' by the operator ``without the source.'' It goes without saying that expecting anything as simple as (\ref{eq:Oscqj}) for the operators of interacting fields would be pointless. As was already noted in section \ref{ch:NBos}, in principle a connection between $\Heis{Q}_j(t)$ and $\Heis{Q}(t)$ is provided by \Eqs{EqUj} and (\ref{eq:QjByQ}). This connection becomes much simpler and more transparent if rewritten in terms of the charactestic functionals. For the oscillator, we observe that functionals $\Phi (\eta _-,\eta _+;j)$ and $\Phi (\eta _-,\eta _+)$ given by (\ref{eq:PhiDef}) and (\ref{eq:PhiDefJ}), respectively, are expressed by the same functional $\Phi \R$, hence there must exist a relation between them. It is found by noting that the replacement \mbox{$\eta \to \eta , \sigma \to \sigma +j$} amounts to \mbox{$\eta _{\pm}\to \eta _{\pm}+ j/\hbar $}, so that 
\begin{gather} 
\begin{aligned}
\Phi (\eta _-,\eta _+;j) = 
\Phi \Big(\eta _- + \frac{j}{\hbar } ,\eta _+ + \frac{j}{\hbar }\Big). 
\end{aligned}%
% \nonumber % 
\eqlabel{SigPlusJ} 
\end{gather}%
Our immediate goal is to prove that this relation holds 
not only for the harmonic oscillator, but also in general for any interacting bosonic system (cf.\ the opening paragraph of section \ref{ch:HarmRevi}). 
We note that \Eq{SigPlusJ} is {\em sine qua non\/} for the very existence of a response formulation of interacting bosons. It makes obvious the fact that $\Heis{Q}(t)$ already contains full information on the response properties of the system. In a sense, the rest of the paper merely clarifies the physical content of \Eq{SigPlusJ}. 

We start the proof of equation (\ref{eq:SigPlusJ}) from interpreting $\Phi (\eta _-,\eta _+;j)$ as Schwinger's amplitude for the evolution forward and backward in time \cite{SchwingerC}: 
\begin{gather} 
\begin{aligned}
\Phi (\eta _-,\eta _+;j) = \qav{\Heis{S}\dg_-\, \Heis{S}_+},  
\end{aligned}%
% \nonumber % 
\eqlabel{SSAv} 
\end{gather}%
where 
$\Heis{S}_{\pm}$ are the forward-in-time and backward-in-time S-matrices, 
\begin{gather} 
\begin{aligned} 
\Heis{S}_{\pm} = {\Heis{U}}_{\pm}(\infty), 
\end{aligned}%
% \nonumber % 
\eqlabel{Spm} 
\end{gather}%
with $\Heis{U}_{\pm}(t)$ being forward-in-time and backward-in-time evolution operators,   
\begin{gather} 
\begin{aligned} 
{\Heis{U}}_{\pm}(t) = T_+\exp \left[
-\frac{i}{\hbar } \int_{-\infty}^t dt'  j_{\pm}(t')\Heis{Q}_j(t')
\right] . 
\end{aligned}%
% \nonumber % 
\eqlabel{TexpUpm} 
\end{gather}%
With 
\begin{gather} 
\begin{aligned}
 j_{\pm}(t) = \hbar \eta _{\pm}(t)    
\end{aligned}%
% \nonumber % 
\eqlabel{EtaToJ} 
\end{gather}%
definition (\ref{eq:SSAv}) obviously coincides with (\ref{eq:PhiDef}), cf.\ also \chKuboSchwinger. For simplicity, in \Eqs{SSAv}--(\ref{eq:TexpUpm}) we assumed that the currents $j_{\pm}$ are real. We firstly verify \Eq{SigPlusJ} for real arguments, then extend this relation to complex $\eta_{\pm}(t)=j_{\pm}(t)/\hbar$. 

Calling ${\Heis{U}}_{\pm}(t)$ {\em evolution operators\/} implies that we can define them as such in some dynamical approach. Proof of (\ref{eq:SigPlusJ}) in fact reduces to finding such approach and carefully writing down all definitions. Namely, consider the following
pair of Hamiltonians in the Schr\"odinger picture,   
\begin{gather} 
\begin{aligned}
\Heis{H}_{\text{S}\pm}(t) = 
\Heis{H}_{0\text{S}}(t) + j(t)\Heis{Q}_{\text{S}} +  j_{\pm}(t)\Heis{Q}_{\text{S}}
. 
\end{aligned}%
% \nonumber % 
\eqlabel{Hpm} 
\end{gather}%
With $ j_{\pm} = 0$ they reduce to the Hamiltonian of the system in \mbox{Fig.\ \ref{fig:RespExp}} given by \Eq{HS}, while $ j_{\pm}(t)\Heis{Q}_{\text{S}}$ has been added 
so as to establish connection with \Eqs{SSAv}--(\ref{eq:EtaToJ}).
Equation (\ref{eq:SigPlusJ}) will be found as a consequence of the trivial fact, that the two constituents of the interaction may be treated either sequentially, or concurrently by combining them into a single term 
\begin{gather} 
\begin{aligned}
j(t)\Heis{Q}_{\text{S}} +  j_{\pm}(t)\Heis{Q}_{\text{S}} = 
 \big[j(t)+ j_{\pm}(t) \big]\Heis{Q}_{\text{S}}. 
\end{aligned}%
% \nonumber % \eqlabel{} 
\end{gather}%
Considering the interaction terms sequentially amounts to representing the evolution operators related to Hamiltonians (\ref{eq:Hpm}) as products of three factors, 
\begin{gather} 
\begin{aligned}
\U_0(t)\U_j(t)\U_{\pm}(t), 
\end{aligned}%
% \nonumber % 
\eqlabel{UUU} 
\end{gather}%
corresponding to the three terms comprising the Hamiltonians (\ref{eq:Hpm}). The factors $\U_0(t)$ and $\U_j(t)$ were introduced in section \ref{ch:NBos}, cf.\ \Eqs{HS}--(\ref{eq:QjByQ}), while 
the operators $\U_{\pm}(t)$ here are the same as given by (\ref{eq:TexpUpm}). 
Indeed, they obey the equations 
\begin{gather} 
\begin{aligned}
i\hbar {\dot {\U}}_{\pm}(t) &=  j_{\pm}(t)\Heis{Q}_j(t) \U_{\pm}(t),  
\end{aligned}%
% \nonumber % 
\eqlabel{UEqs} 
\end{gather}%
solved by (\ref{eq:TexpUpm}). 
Furthermore, considering the interaction terms concurrently means  replacing $\U_j(t)\U_{\pm}(t)$ in (\ref{eq:UUU}) by $\U_{j\pm}(t)$,
\begin{gather} 
\begin{aligned}
\U_{j\pm}(t) = \U_j(t)\U_{\pm}(t).   
\end{aligned}%
% \nonumber % 
\eqlabel{UU} 
\end{gather}%
This pair obey the equations 
\begin{gather} 
\begin{aligned}
i\hbar {\dot {\U}}_{j\pm}(t) = \big[j(t)+ j_{\pm}(t) \big]\Heis{Q}(t) \U_{j\pm}(t), 
\end{aligned}%
% \nonumber % \eqlabel{} 
\end{gather}%
also solved by T-exponents,  
\begin{gather} 
\begin{aligned}
{\Heis{U}}_{j\pm}(t) = T_+\exp \left\{
-\frac{i}{\hbar } \int_{-\infty}^t dt' \big[j(t')+ j_{\pm}(t')\big]\Heis{Q}(t')
\right\} . 
\end{aligned}%
% \nonumber % 
\eqlabel{EqUjpm} 
\end{gather}%
Importantly, the T-exponents (\ref{eq:TexpUpm}) contain $\Heis{Q}_j(t)$ whereas (\ref{eq:EqUjpm}) contain $\Heis{Q}(t)$.   
By making use of (\ref{eq:UU}) then taking $t\to\infty$ we find 
\begin{gather} 
\begin{aligned}
\Heis{S}_{\pm} = \Heis{S}\dg_{j}\, T_+\exp \left\{
-\frac{i}{\hbar } \int dt' \big[j(t')+ j_{\pm}(t')\big]\Heis{Q}(t')
\right\},  
\end{aligned}%
% \nonumber % 
\eqlabel{SpmToSjpm} 
\end{gather}%
where $\Heis{S}_{j} = \Heis{U}_{j}(\infty)$ and $\Heis{S}_{\pm}$ are given by \Eq{Spm}. Equation (\ref{eq:SigPlusJ}) is then recovered by substituting (\ref{eq:SpmToSjpm}) into (\ref{eq:SSAv}), using that 
\begin{gather} 
\begin{aligned}
\Heis{S}_{j}\Heis{S}\dg_{j} = 1 ,   
\end{aligned}%
% \nonumber % 
\eqlabel{SS1} 
\end{gather}%
and recalling (\ref{eq:EtaToJ}). 

To extend these considerations to complex $j_{\pm}$, it suffices to replace $j_-(t)\to j^*_-(t)$ in \Eqs{Hpm}--(\ref{eq:SpmToSjpm}) while preserving all relations involving $j_+(t)$. In particular, we then have
\begin{gather} 
\begin{aligned} 
{\Heis{U}}_{-}(t) = T_+\exp \left[
-\frac{i}{\hbar } \int_{-\infty}^t dt'  
j^*_{-}(t')\Heis{Q}_j(t')
\right] , 
\end{aligned}%
% \nonumber % 
\eqlabel{TexpUmC} 
\end{gather}%
so that 
\begin{gather} 
\begin{aligned} 
{\Heis{U}}\dg_{-}(t) = T_-\exp \left[
\frac{i}{\hbar } \int_{-\infty}^t dt'  j_{-}(t')\Heis{Q}_j(t')
\right] , 
\end{aligned}%
% \nonumber % 
\eqlabel{TexpUCm} 
\end{gather}%
leaving equation (\ref{eq:SSAv}) unchanged. Importantly, the Kubo current remains real, so that none of the physical quantities occuring in \Eqs{Hpm}--(\ref{eq:SpmToSjpm}) are affected; the only affected ones are the evolution operators $\Heis{U}_{\pm}(t)$ and $\Heis{U}_{j\pm}(t)$ which are anyway purely formal. Equation (\ref{eq:SigPlusJ}) is thus also proven for complex $\eta_{\pm}(t)=j_{\pm}(t)/\hbar$. 
%******************************************* 
\section{Response formulation of interacting bosons}\chlabel{ch:PhysPr}
%******************************************* 
%******************************************* 
\subsection{Preliminary considerations}\chlabel{ch:Prelim}
%******************************************* 
Response reformulation of interacting bosons is a straigtforward generalisation of \Eq{qjnorm} for the harmonic oscillator. We simply replace the operator $\hat q_j(t)$ by the Heisenberg field operator $\Heis{Q}_j(t)$, and the normal ordering of the free-field operators by the {\em time-normal\/} ordering of Heisenberg operators. The latter was introduced by Kelly and Kleiner \cite{KellyKleiner} in the context of macroscopic photodetection, see also \cite{GlauberTN,MandelWolf,Schleich,Corresp}. 
The ``interacting'' generalisation of (\ref{eq:qjnorm}) then reads
\begin{multline} % 
\preprintmargin
\Phi_{\text{R}}(\eta;j) = \qav{{\mathcal T}\bm{:}\exp\int dt\, \eta (t)\Heis{Q}_j(t)\bm{:}}
\\ = 
1 + \sum_{m=1}^{\infty} \frac{1}{m!} 
\left(
\int dt 
\right)^m 
\eta (t_1)\cdots\eta (t_m) 
%\\ \times 
\qav{{\mathcal T}\bm{:}\Heis{Q}_j(t_1)\cdots\Heis{Q}_j(t_m)\bm{:}} 
, 
\eqlabel{PhiR} % 
\preprintmargin
\end{multline}% 
where the notation $ {\mathcal T}
\bm{:}
\cdots
\bm{:} $ for the time-normal ordering is borrowed from Mandel and Wolf \cite{MandelWolf}. 
Within the approximation of slowly varying amplitudes, the time-normal averages in (\ref{eq:PhiR}) may be calculated by expanding each operator into the  
frequency-positive and negative parts, 
\begin{gather} 
\begin{aligned}
\Heis{Q}_j(t) = \Heis{Q}_j^{(+)}(t) + \Heis{Q}_j^{(-)}(t) ,  
\end{aligned}%
% \nonumber % 
\eqlabel{ToPM} 
\end{gather}%
then applying the Glauber-Kelly-Kleiner (GKK) definition, 
%=============================================
\eqM{ 
{\mathcal T}\bm{:} 
\Heis{Q}_j^{(-)}(t _1)\cdots\Heis{Q}_j^{(-)}(t _{\bar m})
\Heis{Q}_j^{(+)}(t _{\bar m+1})\cdots\Heis{Q}_j^{(+)}(t _m) 
\bm{:}
\\ =  
T_-\Heis{Q}_j^{(-)}(t _1)\cdots\Heis{Q}_j^{(-)}(t _{\bar m})
T_+\Heis{Q}_j^{(+)}(t _{\bar m+1})\cdots\Heis{Q}_j^{(+)}(t _m) .  
}{
\eqlabel{TNQO} 
}% 
%+++++++++++++++++++++++++++++++++++++++++++++
The reason why the GKK definition cannot be used in general will be made clear in the next paragraph, and an amended definition will be given in section \ref{ch:RedefTN}. Discussion of this amended definition will, to a large extent, be subject of the rest of the paper.  

Physics behind (\ref{eq:PhiR}) is best understood if we consider the alternative way of introducing the response formulation: by analogy with the decription of 
the Gedanken experiment in Fig.\ \ref{fig:RespExp} in the classical 
stochastic field theory. 
As was mentioned in section \ref{ch:HarmRevi}, the quantum and classical response formulations of the harmonic oscillator are structurally identical. 
They differ only in the replacement of the classical stochastic averages by the normal averages. We now show that \Eq{PhiR} follows if we replace the stochastic averages in the classical description of the experiment in Fig.\ \ref{fig:RespExp} by the time-normal averages. This means, in particular, that the structural identity between the classical and quantum response pictures is a general property of bosonic systems. 

In the most general classical approach, 
the field seen by the detector in Fig.\ \ref{fig:RespExp} is 
a c-number stochatic field 
${\mathcal Q}_j(t)$ dependent on the external source $j(t)$. 
It is fully characterised by its stochastic moments    
$\av{{\mathcal Q}_j(t_1)\cdots{\mathcal Q}_j(t_m)}$ .   
In turn, the field response is fully characterised by the 
classical stochastic response functions
%=============================================
\eqA{ 
{{\mathcal D}}^{(m,n)}_{\text{C}}
(t_1,\cdots,t_m;t'_1,\cdots,t'_n)
=  %\\ 
\frac{\delta^n\av{{\mathcal Q}_j(t_1)\cdots{\mathcal Q}_j(t_m)}
}{\delta j(t'_1)\cdots\delta j(t'_n)} \Big |_{j=0}\, .  
}{
\eqlabel{DCDef} % 
% \nonumber % \eqlabel{} 
}% 
%+++++++++++++++++++++++++++++++++++++++++++++
As was first shown by Glauber \cite{GlauberPhDet},  
and later extended to Heisenberg fields by 
Kelly and Kleiner \cite{KellyKleiner} (see also \cite{GlauberTN,MandelWolf,Schleich,Corresp}), the quantum generalisation of this 
classical picture consists of replacing the classical stochastic averages by the time-normal averages of the quantum field 
\begin{gather} 
\begin{aligned}
\av{{\mathcal Q}_j(t_1)\cdots{\mathcal Q}_j(t_m)} 
\to 
\qav{{\mathcal T}\bm{:}\Heis{Q}_j(t_1)\cdots\Heis{Q}_j(t_m)\bm{:}} 
. 
\end{aligned}%
% \nonumber % 
\eqlabel{Cl2TN} 
\end{gather}%
The response of the quantum field is then characterised 
by the {\em quantum-statistical 
response functions\/}
\begin{multline} % 
\preprintmargin
{{\mathcal D}}^{(m,n)}_{\text{R}}
(t_1,\cdots,t_m;t'_1,\cdots,t'_n)
= %\\  
\frac{\delta^n\qav{{\mathcal T}\bm{:}\Heis{Q}_j(t_1)\cdots\Heis{Q}_j(t_m)\bm{:}}}
{\delta j(t'_1)\cdots\delta j(t'_n)}  |_{j=0} 
\\ =  
\frac{\delta^{n}}
{ 
\delta j(t'_1)\cdots\delta j(t'_n)
}\left.\left[
\frac{\delta^{m}\Phi_{\text{R}}(\eta;j)}
{\delta\eta (t_1)\cdots\delta\eta (t_m) }
|_{\eta  =0}  
\right]\right| 
_{j =0}  
,  
\eqlabel{DRDef} % 
\preprintmargin
\end{multline}% 
where $\Phi_{\text{R}}(\eta;j)$ is the characteristic functional of the time-normal averages of $\Heis{Q}_j(t)$ given by (\ref{eq:PhiR}).

In addition to clarifying the physical content of the definition (\ref{eq:PhiR}), \Eqs{DRDef} show that interpetation of the functional $\Phi\R (\eta ;j )$ is 
by construction two-fold. 
Viewed as a functional of $\eta $ with $j$ a parameter, 
$\Phi\R (\eta ; j )$ is a characteristic functional 
of the time-normal averages of operator $\Heis{Q}_j(t)$. 
Viewed as a functional of two variables, 
$\Phi\R (\eta ; j )$ is a characteristic functional 
of the quantum-statistical response functions of the quantum field,  
%%\begin{widetext}%
\begin{multline} % 
\preprintmargin
\Phi\R (\eta ; j ) = 1 + \sum_{m+n\geq 1} \frac{1}{m!n!} 
%\,\\ \times 
\left(
\int dt
\right)^{m+n} 
\eta (t_1)\cdots\eta (t_m) 
j   (t'_1)\cdots j  (t'_n) 
\,\\ \times 
{{\mathcal D}}^{(m,n)}_{\text{R}}
(t_1,\cdots,t_m;t'_1,\cdots,t'_n)
, 
\eqlabel{PhiRD} 
\preprintmargin
\end{multline}% 
%%\end{widetext}%
where $\big(
\int dt
\big)^{m+n}$ stands for  integration over 
all  $t$s, cf.\ also endnote \cite{IntRange}. 
Consistency of these two 
interpretations of $\Phi\R (\eta ;j)$ is warranted 
by equations (\ref{eq:DRDef}). 
%******************************************* 
\subsection{Failure of the naive generalisation}
%******************************************* 
A discouraging observation is that the functional $\Phi \R(\eta ;j)$ defined by (\ref{eq:PhiR})--(\ref{eq:TNQO}) cannot be mapped exactly on $\Phi (\eta _-,\eta _+;j)$ given by (\ref{eq:PhiDefJ}). 
Consider, e.g., the time-normal average of three field operators 
$\tnav{\Heis{Q}_j(t_1)\Heis{Q}_j(t_2)\Heis{Q}_j(t_3)}$. 
Separation of the frequency-positive and negative 
parts of a function is conveniently expressed as an integral operation,  
\begin{gather} 
g^{(\pm)}(t) = 
\int dt' \delta^{(\pm)}(t-t') g(t')  , 
\eqlabel{FPNAsConv} 
\end{gather}%
where $\delta^{(\pm)}(t)$ are the frequency-positive and negative parts of the $\delta$-function, cf.\ appendix \ref{app:FPN}. 
On assuming that $t_1>t_2>t_3$ and pairwise combing 
terms we find 
%%\begin{widetext} 
\begin{multline} % 
{\mathcal T}  
\bm{:}\Heis{Q}_j(t_1)\Heis{Q}_j(t_2)\Heis{Q}_j(t_3)\bm{:}  = %\\ 
\left(
\int dt'
\right) ^3 \Heis{Q}_j(t'_1)\Heis{Q}_j(t'_2)\Heis{Q}_j(t'_3) 
\\ \times %\hspace{0.2\columnwidth} \\ 
\Big[
\delta^{(-)}(t_3-t'_1)\delta^{(-)}(t_2-t'_2)\delta(t_1-t'_3) 
+  
\delta^{(-)}(t_2-t'_1)\delta(t_1-t'_2)\delta^{(+)}(t_3-t'_3) 
\\ + 
\delta^{(-)}(t_3-t'_1)\delta(t_1-t'_2)\delta^{(+)}(t_2-t'_3) 
+  
\delta(t_1-t'_1)\delta^{(+)}(t_2-t'_2)\delta^{(-)}(t_3-t'_3) 
\Big] 
. 
\eqlabel{GKK3} 
\end{multline}% 
%%\end{widetext}%
The integration here is over all possible values of $t'_1,t_2',t_3'$ 
including $t_2'<t_1',t_3'$. 
However, for this order of arguments, the product $\Heis{Q}_j(t'_1)\Heis{Q}_j(t'_2)\Heis{Q}_j(t'_3)$ is not double-time ordered. Indeed, in a double-time-ordered product, the times are arranged into a first-increase-then-decrease sequence, which 
\mbox{$t_1'>t_2'<t_3'$} is obviously not. 
One result of paper \cite{RespOsc} is that for the free-field operators such offending contributions cancel. Unlike the normal products of the free-field operators, the time-normal products of the Heisenberg operators as defined by Glauber and Kelly and Kleiner cannot be expressed solely by the double-time-ordered products of the field operators. 
%******************************************* 
\subsection{Redefining the time-normal ordering}\chlabel{ch:RedefTN}
%******************************************* 
To find a way of dealing with the complication we have just encountered, let us have another look at the results for the harmonic oscillator. Combining equations (\ref{eq:qjnorm}) and (\ref{eq:PhiJbyPhiR}) we find 
\begin{multline} % 
\preprintmargin
\bm{:}\exp\int dt\, \eta (t)
\hat q_j (t)\bm{:} \\ =  
T_-\exp \int dt\, \eta^{(+)}(t)\hat q_j(t)\, %\\ \times     
T_+\exp \int dt\, \eta^{(-)}(t)\hat q_j(t) 
. 
\eqlabel{NDef} % 
\preprintmargin
\end{multline}% 
Dropping the quantum averaging here is justified, because \Eqs{qjnorm} and (\ref{eq:PhiJbyPhiR}) hold for an arbitrary quantum state. The LHS of (\ref{eq:NDef}) is an operator-valued characteristic functional of the normal products of $\hat q_j(t)$, so that this relation may be used as an alternative definition of the normal ordering. 
We follow this pattern and {\em redefine\/} the time-normal ordering of Heisenberg operators as  
\begin{multline} % 
\preprintmargin
{\mathcal T}\bm{:}\exp\int dt\, \eta (t)
\Heis{Q}_j (t)\bm{:} \\ =  
T_-\exp \int dt\, \eta^{(+)}(t)\Heis{Q}_j(t)\, %\\ \times     
T_+\exp \int dt\, \eta^{(-)}(t)\Heis{Q}_j(t) 
.  
\eqlabel{TNDef}  % 
\preprintmargin
\end{multline}% 
%%\end{widetext}%
By virtue of shorthand (\ref{eq:xY}), this defines the time-normal products not just for a nonlinear oscillator, but for an arbitrary bosonic field. In the field case, the positive and negative-frequency parts 
always apply to the time variable. 

Unlike (\ref{eq:NDef}), \Eq{TNDef} is not equivalent to the conventional definition by Glauber and Kelly and Kleiner (GKK definition). It is therefore instructive to understand the difference between definitions (\ref{eq:TNQO}) and (\ref{eq:TNDef}). There is a very simple {\em incorrect\/} way of deriving the GKK definition from \Eq{TNDef}, by using the formula 
\eqA{
\int dt \eta^{(\pm)}(t)\Heis{Q}_j(t) = 
\int dt\eta(t)\Heis{Q}_j^{(\mp)}(t). 
}{\eqlabel{ShiftToQ}}% 
This formula follows from \Eq{FPNAsConv} and from 
%=============================================
\eqA{ 
\delta^{(\pm)}(t-t') = 
\delta^{(\mp)}(t'-t) ,
}{
% \nonumber % \eqlabel{} 
}% 
%+++++++++++++++++++++++++++++++++++++++++++++
cf.\ \Eq{DeltaPMInv} in the appendix. Then, indeed, 
%=============================================
\eqA{ 
\int dt \eta^{(\pm)}(t)\Heis{Q}_j(t) = %\\ 
\int dt dt' \eta(t)\delta^{(\mp)}(t-t')\Heis{Q}_j(t') =  
\int dt'\eta(t')\Heis{Q}_j^{(\mp)}(t'). 
}{\eqlabel{ShiftToQ0}}% 
%+++++++++++++++++++++++++++++++++++++++++++++
Applying \Eq{ShiftToQ} to (\ref{eq:TNDef}) we have, 
\eqM{
{\mathcal T}\bm{:}\exp\int dt\, \eta (t)
\Heis{Q}_j (t)\bm{:} \\ =  
T_-\exp \int dt\, \eta(t)\Heis{Q}^{(-)}_j(t)\, %\\ \times     
T_+\exp \int dt\, \eta(t)\Heis{Q}^{(+)}_j(t) 
.  
}{\eqlabel{TNDefFl}}% 
It is easy to check that this formula is equivalent to the GKK definition (\ref{eq:TNQO}).

To understand the flaw in this derivation is to understand the difference between (\ref{eq:TNQO}) and (\ref{eq:TNDef}). Consider a term in (\ref{eq:TNDef}) contributing to ${\mathcal T}\bm{:}
\Heis{Q}_j (t)
\Heis{Q}_j (t')\bm{:}$, 
\eqM{
\int dt dt' \eta^{(-)}(t)\eta^{(-)}(t')T_+\Heis{Q}_j(t)\Heis{Q}_j(t') \\ =  
\int dt dt'\eta(t) \eta(t')\int  dt'' dt''' \delta^{(+)}(t-t'')
%\,\\ \times  \delta^{(+)}(t'-t''') 
T_+\Heis{Q}_j(t'')\Heis{Q}_j(t'''), 
}{\eqlabel{TNQQpp}}% 
where use was made of \Eq{ShiftToQ0}. The inner integral in (\ref{eq:TNQQpp}) expresses separation of the frequency-positive parts of $T_+\Heis{Q}_j(t'')\Heis{Q}_j(t''')$ in respect of $t''$ and $t'''$, cf.\ \Eq{FPNAsConv}. In \Eq{TNQQpp} (and also in general in (\ref{eq:TNDef})) the time orderings come first, and the separation of the frequency-positive and negative parts second. In the GKK definition, separation of the frequency-positive and negative parts comes first, and time orderings second. 
The flaw in the above derivation is that it assumes implicitly that these operations commute. In fact they do not, and this is exactly how the difference between (\ref{eq:TNQO}) and (\ref{eq:TNDef}) originates.

It is convenient to preserve the customary notation for the time-normal operator products by introducing the shorthand   
%=============================================
%\begin{widetext} 
\eqA{ 
T_{\pm} \big[\cdots\Heis{Q}_j^{(\pm)}(t)\cdots\big]  
=  % \\ 
\int dt' \delta^{(\pm)} (t-t') 
T_{\pm} \big[\cdots\Heis{Q}_j(t')\cdots\big]   
.  
}{
% \nonumber % 
\eqlabel{Shorthand} 
}% 
%+++++++++++++++++++++++++++++++++++++++++++++
Using this shorthand means that the GKK definition (\ref{eq:TNQO}) has been replaced by 
\begin{multline} % \preprintmargin
{\mathcal T}\bm{:} 
\Heis{Q}_j^{(-)}(t _1)\cdots\Heis{Q}_j^{(-)}(t _{\bar m})
\Heis{Q}_j^{(+)}(t _{\bar m+1})\cdots\Heis{Q}_j^{(+)}(t _m) 
\bm{:}
\\ =  
\left(
\int dt'
\right)^m 
\delta^{(-)}(t_1-t_1')
\cdots 
\delta^{(-)}(t_{\bar m}-t_{\bar m}') 
\delta^{(+)}(t_{\bar m+1}-t_{\bar m+1}')
\cdots 
\delta^{(+)}(t_m-t_m')
\\ \times  
T_-\Heis{Q}_j(t' _1)\cdots\Heis{Q}_j(t '_{\bar m})
T_+\Heis{Q}_j(t '_{\bar m+1})\cdots\Heis{Q}_j(t' _m)
.     
\eqlabel{TNGen} % \preprintmargin
\end{multline}% 
The order of the separation of the frequency-positive and 
negative parts and of the time-ordering here is obviously reversed compared to (\ref{eq:TNQO}).

Usefulness of shorthand (\ref{eq:Shorthand}) is in fact twofold. Firstly, expressions involving this shorthand may be handled algebraically exactly as those with the GKK definition. For instance, the formulae 

%=============================================
\eqA{ 
& \formA{ 
{\mathcal T}\bm{:}\Heis{Q}_j(t)\bm{:} \A = 
{\mathcal T}\bm{:}
\Heis{Q}_j^{(+)}(t)+\Heis{Q}_j^{(-)}(t)
\bm{:} \\ \A = 
T_+\Heis{Q}_j^{(+)}(t) +
T_-\Heis{Q}_j^{(-)}(t) = 
\Heis{Q}_j^{(+)}(t) + \Heis{Q}_j^{(-)}(t) = \Heis{Q}_j(t), 
} % \nonumber % \eqlabel{} 
\\ & \formA{ 
{\mathcal T}\bm{:}\Heis{Q}_j(t)\Heis{Q}_j(t')\bm{:} \A = 
{\mathcal T}\bm{:}
\Sbracket{\big}{
\Heis{Q}_j^{(+)}(t)+\Heis{Q}_j^{(-)}(t)
}\Sbracket{\big}{
\Heis{Q}_j^{(+)}(t')+\Heis{Q}_j^{(-)}(t')
} \bm{:} \\ \A = % \hspace{0.2\columnwidth}  
T_+\Heis{Q}_j^{(+)}(t)\Heis{Q}_j^{(+)}(t') +
\Heis{Q}_j^{(-)}(t)\Heis{Q}_j^{(+)}(t')  
\\ \A \ \ \ + %\hspace{0.4\columnwidth} 
\Heis{Q}_j^{(-)}(t')\Heis{Q}_j^{(+)}(t) + 
T_-\Heis{Q}_j^{(-)}(t)\Heis{Q}_j^{(-)}(t'), 
} % \nonumber % \eqlabel{} 
}{
% \nonumber % 
\eqlabel{AlgEx} % \A 
}% 
%+++++++++++++++++++++++++++++++++++++++++++++
%\end{widetext}% 
etc., hold with either definition. Secondly, and more importantly, definitions (\ref{eq:TNQO}) and (\ref{eq:TNGen}) need not be distinguished within the quantum-optical paradigm. 
Indeed, analyses in \cite{RespOsc}  
verify that, for the free fields, definitions (\ref{eq:TNQO}) and (\ref{eq:TNGen}) are exactly equivalent, so that there should exist physical conditions under which these definitions remain approximately equivalent. It is easy to see that one such condition is the approximation of the slowly varying amplitudes. As a consequence, our redefinition of the time-normal ordering does not affect interpretation of any of the experimental facts in quantum optics involving this concept.  
The reader who is only intererested in applying  our results 
within the conventional quantum-optical paradigm may ignore our 
specifications. 
The reader who is interested in how the definitions 
initially developed within this paradigm  
must be changed in order to apply to relativistic 
quantum fields (say) should remember that here, strictly speaking,   
(\ref{eq:TNQO}) is just a shorthand for (\ref{eq:TNGen}). 
%******************************************* 
\subsection{Response reformulation of bosons}
%******************************************* 
The response formulation of interacting bosons is {\em by definition\/}  introduced by \Eqs{PhiR}, (\ref{eq:DRDef}) and (\ref{eq:PhiRD}), where the symbol of the time-normal ordering must be understood according to (\ref{eq:TNDef}) in place of (\ref{eq:TNQO}).
The immediate consequence of this definition is that the fundamental relation (\ref{eq:PhiJbyPhiR}) connecting the closed-time-loop and the response formalisms holds also for the interacting bosons. 
Indeed, on averaging \Eq{TNDef} and applying definitions (\ref{eq:PhiDefJ}) and (\ref{eq:PhiR}) we find
\eqA{
\Phi \R(\eta ,j) = \Phi (-i\eta ^{(+)},i\eta ^{(-)};j) 
%\\ =  
%\Phi \left(
%-i\eta ^{(+)} + \frac{\sigma }{\hbar } ,
%i\eta ^{(-)}+ \frac{\sigma }{\hbar } 
%\right) 
.  
}{
\eqlabel{FromTN} % 
}%  
It is easy to see that \Eq{SigPlusJ} affords a trivial generalisation, namely, for any $j_1$ and $j_2$, 
\eqA{
\Phi(\eta_-,\eta_+;j_1+j_2) = 
\Phi\Big(\eta_- +\frac{j_1}{\hbar},\eta_+ +\frac{j_1}{\hbar};j_2\Big). 
}{\eqlabel{j1j2}}% 
This relation is verified by applying \Eq{SigPlusJ} to both sides of it. 
Replacing $j\to j+\sigma$ in \Eq{FromTN} and using \Eq{j1j2} then results in 
%=============================================
\eqA{ 
\Phi \R(\eta ,\sigma + j) = %\\ 
\Phi \left(
-i\eta ^{(+)} + \frac{\sigma }{\hbar } ,
i\eta ^{(-)}+ \frac{\sigma }{\hbar } ; j 
\right) = \Phi \left(
\eta_-,
\eta_+; j 
\right) , 
}{
\eqlabel{PhiJR0} % 
% \nonumber % \eqlabel{} 
}% 
%+++++++++++++++++++++++++++++++++++++++++++++
where the pairs of functional variables $\eta^{\mu }_{\pm}(\bm{r},t)$ and $\eta^{\mu } (\bm{r},t),\sigma^{\mu }(\bm{r},t)$ are coupled by the substitutions 
\begin{gather} 
\begin{aligned}
\eta^{\mu } (\bm{r},t) & = - i 
\left[\eta^{\mu } _+(\bm{r},t)-\eta^{\mu } _-(\bm{r},t)\right], \\  
\sigma^{\mu }(\bm{r},t) & = \hbar \left[
\eta^{\mu(+)} _+(\bm{r},t) + \eta^{\mu(-)} _-(\bm{r},t)
\right] , 
\end{aligned}%
% \nonumber % 
\eqlabel{SubstEta} 
\\ 
\begin{aligned}
\eta^{\mu }_+(\bm{r},t) & = 
i \eta^{\mu(-)} (\bm{r},t) + 
\frac{1}{\hbar} \sigma^{\mu }(\bm{r},t), \\    
\eta^{\mu }_-(\bm{r},t) & = - i\eta^{\mu(+)} (\bm{r},t) + 
\frac{1}{\hbar} \sigma^{\mu }(\bm{r},t) .   
\end{aligned}%
% \nonumber % 
\eqlabel{SubstSigma} 
\end{gather}%
We have thus recovered relation (\ref{eq:PhiJbyPhiR}) for interacting fields, with substitutions (\ref{eq:OscSubstEta}), (\ref{eq:OscSubstSigma}) replaced by their field versions. Substitutions (\ref{eq:SubstEta}), (\ref{eq:SubstSigma}) are exactly those found in \chFieldN\ for noninteracting bosonic fields ensuring overall consistency. Note that applying \Eq{j1j2} in deriving \Eq{PhiJR0} implies that $\sigma^{\mu }(\bm{r},t)$ is real. 
This assumption is not an obstacle, cf.\ the discussion \DiscEndSecII. We return to this condition shortly.

The logic of the above argument may be inverted by using substitutions  (\ref{eq:SubstEta}) and (\ref{eq:SubstSigma}) as a starting point in place of \Eq{TNDef}. 
The functional $\Phi\R(\eta;j)$ is then defined {\em a priori\/} by applying (\ref{eq:SubstSigma}) to $\Phi(\eta_-,\eta_+)$, 
\begin{align} 
\Phi\R (\eta ;\sigma ) & = \Phi (\eta _-,\eta _+)|_{
\eta_{\pm} = \hbar ^{-1}\sigma \pm i\eta^{(\mp)}
} ,  
\eqlabel{ToPhiR} 
\end{align}% 
while equations (\ref{eq:PhiR}), (\ref{eq:DRDef}) and (\ref{eq:PhiRD}) serve as equally {\em a priori\/} definitions of the time-normal averages and response functions (independent of the GKK definitions). Equation (\ref{eq:PhiJR0}) becomes a consequence of (\ref{eq:j1j2}) and (\ref{eq:ToPhiR}), while \Eq{FromTN} is found as a particular case of (\ref{eq:PhiJR0}) with $\sigma = 0$. The formula for the time-normal products (\ref{eq:TNDef}) is then recovered from (\ref{eq:FromTN}) by noting that the latter must hold for an arbitrary state of the field. Importantly, irrespective of whether we start from \Eq{TNDef} or from \Eq{ToPhiR}, the overall consistency of our definitions depends on  \Eq{SigPlusJ}.   

In essence, 
we have generalised the results for the oscillator to quantum fields by postulating that 
substitutions (\ref{eq:OscSubstEta}) and (\ref{eq:OscSubstSigma}) 
are always applied to the time argument and do not touch field labels. 
This way of extending the results for the 
oscillator to fields avoids the concept of mode, making our considerations applicable, in particular, in the case 
of arbitrarily strong interactions, when introducing modes may be difficult or impossible. 

Despite the generality of our definitions, we will be able to 
satisfy explicit causality in the response formulation and trace its links to Glauber's photodetection theory and Kubo's linear response theory. 
In particular, we shall show that all reality and causality properties 
characteristic of the classical stochastic response functions 
(\ref{eq:DCDef}) also hold for the quantum-statistical response 
functions (\ref{eq:DRDef}). 
It is easy to see that the time-normal averages introduced by 
definition (\ref{eq:ToPhiR}) are real quantities; the ${{\mathcal D}}^{(m,n)}_{\text{R}}$s are therefore also real. 
This becomes obvious if $\eta ,\sigma $ are chosen real; this choice was in fact already implied when deriving \Eq{PhiJR0}. 
Then $\eta _- = \eta _+^*$ making  
$\Phi\R (\eta ;\sigma) = \Phi (\eta _-,\eta _+)|_{
\eta_{\pm} = \hbar ^{-1}\sigma \pm i\eta^{(\mp)}}$ real. 
Coefficients of the Taylor expansion of a real functional, cf.\ \Eq{PhiR}, 
are obviously real quantities. 
Causality in the response formulation will be demonstrated in section \ref{ch:Caus}.

It cannot be emphasised strongly enough that 
equations (\ref{eq:TNDef})--(\ref{eq:ToPhiR})  
constitute a radical deviation from the logic of leading considerations in section \ref{ch:Prelim}. 
Equation (\ref{eq:DRDef}) implies that the concept of time-normal ordering is known, and uses it to define the response functions, 
whereas equation (\ref{eq:ToPhiR}) 
postulates the response formulation thus {\em redefining\/} the  time-normal ordering. 
This change of the logic, we remind the reader, derives from the fact that the response formulation based on the conventional 
time-normal ordering as defined by Glauber and Kelly and Kleiner \cite{KellyKleiner,GlauberTN} does not map onto the Green-function formulation. 
This forces us to consider \Eq{ToPhiR} as an {\em a priori definition\/}, 
and the normal averages and the quantum-statistical 
response functions---as {\em purely structural 
objects\/} defined by expanding functional $\Phi\R (\eta ;\sigma )$ 
in functional Taylor series (\ref{eq:PhiR}) and (\ref{eq:PhiRD}), 
respectively. 

At the same time, response formulation based on definition (\ref{eq:TNDef}) is by construction equivalent to the closed-time-loop formalism. 
%We believe that this alone suffices to justify our redefinition. 
The quantum-statistical response functions are then expressed 
by Green's functions (\ref{eq:FGNL}), and, importantly, Green's functions 
are expressed 
by the quantum-statistical response functions. 
In less formal words, everything about 
the quantum system is in its self-radiation and response, 
and that 
nothing exists that is decribed by quantum mechanics 
and that cannot be 
reformulated in self-radiation and/or response terms. 
As has already been mentioned in the introduction, 
it also shows that the information contained in the field operator 
(and in the intial condition)
suffices to describe any {\em scattering experiment\/} performed on 
the system. It is worthy of stressing that all these results hinge on \Eq{SigPlusJ}. While by itself this formula is in no way limited to the response formulation, this formulation certainly makes its physical interpretation much more transparent. 
%******************************************* 
%\newpage
\section{Structural response properties of interacting bosonic fields%
}\chlabel{ch:NOscResp} 
%******************************************* 
\subsection{The linear response}\chlabel{ch:LinResp} 
%******************************************* 
First indication that the way we introduced the response functions makes 
physical sense comes from considering the linear response function ${\mathcal D}\R^{(1;1)}(t;t')$. 
It is calculated directly from definitions (\ref{eq:PhiDef}), 
(\ref{eq:DRDef}) and (\ref{eq:ToPhiR}). 
After some straighforward algebra we obtain  

\begin{gather} 
{\mathcal D}_{\text{R}} ^{(1;1)}(t;t') 
= -\frac{i}{\hbar } 
\theta(t - t') \qav{\Big[\Heis{Q}(t),
{\Heis{Q}}(t')\Big]}  
,  
\eqlabel{FieldKuboA} 
\end{gather}%
which is nothing but Kubo's renown formula \cite{Kubo} for the linear 
response function. 
Details of its derivation based on  
the general formula for ${\mathcal D}\R^{(m,n)}$ 
are given in section \ref{ch:CausKubo}. 
Recalling that our definition of the response functions 
hinges on a structural analogy with a linear system, 
rediscovering Kubo's formula which holds for any 
nonlinear system is quite encouraging. 
We note also that Kubo's formula is the simplest 
example of the general link between commutators and response 
to be encountered throughout this paper. 
%******************************************* 
\subsection{Quantum response and noncommutativity of operators}\chlabel{ch:RespToSourceFunc} 
%******************************************* 
A closer inspection of \Eqs{SigPlusJ} and (\ref{eq:PhiJR0}) 
reveals a fundamental 
link between the operator noncommutativity and the quantum-statictical 
response of the quantum system. 
Indeed, if we assume that $\Heis{Q}(t)$ for different 
times commute with 
each other, the time orderings in definitions of $\Phi (\eta _-,\eta _+;j)$ 
and $\Phi (\eta _-,\eta _+)$ may be dropped so we have 
%=============================================
\eqA{ 
\Phi (\eta _-,\eta _+;j) = %\\ 
\bigg\langle
\exp 
\Big\{-i\int dt\big[ 
\eta _+(t) -  \eta _-(t) 
\big] {\hat{{\mathcal Q}}}_j(t) \Big\} 
\bigg \rangle \equiv F(\eta ;j) , 
}{
\eqlabel{Fj} 
% \nonumber % \eqlabel{} 
}% 
%+++++++++++++++++++++++++++++++++++++++++++++
and 
%=============================================
\eqA{ 
\Phi (\eta _-,\eta _+) = %\\ 
\bigg\langle
\exp 
\Big\{-i\int dt\big[ 
\eta _+(t) -  \eta _-(t) 
\big] {\hat{{\mathcal Q}}}(t) \Big \} 
\bigg\rangle \equiv F(\eta ) , 
}{
\eqlabel{F} 
% \nonumber % \eqlabel{} 
}% 
%+++++++++++++++++++++++++++++++++++++++++++++
while (\ref{eq:SigPlusJ}) shows that 
\begin{gather} 
\begin{aligned}
F(\eta ;j) = F(\eta ).  
\end{aligned}%
% \nonumber % 
\eqlabel{FjEqF} 
\end{gather}%
Neglecting the noncommutativity of operators thus fully cancels the dependence 
on the external source. We further discuss this point 
in the following two paragraphs. 
%******************************************* 
\subsection{General formula for nonlinear 
quantum-statictical response functions}\chlabel{ch:FullResp} 
%******************************************* 
The question we address now is how the response functions of the 
field are expressed by the averages of operator $\Heis{Q}(t)$. 
We note immediately that the very fact that such a relation 
exists is highly nontrivial. 
Moreover, this relation must be inherently quantum. 
As we demonstrate now, in the quantum field theory the response infomation is contained in the commutators. In other words, commutators of a Heisenberg field express formally the reaction of the field to perturbations. In the classical field theory, the c-number fields commute so that this information must be expressed by other means: it enters through the Hamilton-Jacobi equations or some other dynamical relations. 

A formula for 
${\mathcal D}^{(m;n)}$ follows by expanding the 
RHS of \Eq{ToPhiR} into power series and comparing them to series 
(\ref{eq:PhiRD}). 
Expansion of the 
RHS of \Eq{ToPhiR} is given by \Eq{PhiDef}; by using it we have 
%\begin{widetext} 
%=============================================
\eqMW{ 
\Phi \R(\eta;j) = \Phi (\eta _-,\eta _+)|_{
\eta_{\pm} = \hbar ^{-1}j\pm i\eta^{(\mp)}} = %\\ 
1+  
\sum_{k+l\geq 1} \frac{i^k(-i)^l}{k!\,l!} \, \\ \times  
\Qbracket{\Bigg}{ 
T_- 
\Cbracket{\bigg}{
\int dt 
\Sbracket{\bigg}{
\frac{j(t)}{\hbar } - i\eta \FP(t)
}\Heis{Q}(t) 
}^k \, % \\ \times  
T_+ 
\Cbracket{\bigg}{
\int dt 
\Sbracket{\bigg}{
\frac{j(t)}{\hbar } + i\eta \FM(t)
}\Heis{Q}(t) 
}^l 
} . 
}{
% \nonumber % 
\eqlabel{GenDmnXX} 
}% 
%+++++++++++++++++++++++++++++++++++++++++++++
Applying the binomial expansion we have 
%=============================================
\eqMW{ 
\Phi \R(\eta;j) = 1 + 
\sum_{k+ l\geq 1} 
\sum_{\bar k=0}^k \sum_{\bar l=0}^l 
\frac{i^{k+\bar l}(-i)^{l+\bar k}}
{\bar k!\,(k-\bar k)!\,\bar l! \,(l-\bar l)!\,
\hbar ^{k+l-\bar k - \bar l}} \, \\ \times  
\Qbracket{\Bigg}{ 
T_- 
\Sbracket{\bigg}{
\int dt \eta\FP (t)\Heis{Q}(t) 
}^{\bar k}
\Sbracket{\bigg}{
\int dt j(t)\Heis{Q}(t) 
}^{k-\bar k}
 \, \\ \times  T_+ 
\Sbracket{\bigg}{
\int dt \eta\FM (t)\Heis{Q}(t) 
}^{\bar l}
\Sbracket{\bigg}{
\int dt j(t)\Heis{Q}(t) 
}^{l-\bar l}
\,} 
. 
}{
% \nonumber % \eqlabel{} 
}% 
%+++++++++++++++++++++++++++++++++++++++++++++
Comparing this to (\ref{eq:PhiRD}) we see that the terms contributing to $D\R^{(m;n)}$ are those with $\bar k+\bar l = m$, $k+l -\bar k-\bar l=n$. It is therefore convenient to introduce new summation variables such that 
%=============================================
\eqA{ 
\bar k \A = \bar m, \A
k - \bar k \A= \bar n,\A
\bar l \A= m - \bar m, \A
l-\bar l  \A= n- \bar n
, 
}{
% \nonumber % \eqlabel{} 
}% 
%+++++++++++++++++++++++++++++++++++++++++++++
and 
%=============================================
\eqMW{ 
\Phi \R(\eta;j) = 1+  
\sum_{m+n\geq 1} \sum_{{\bar m}=0}^m \sum_{{\bar n}=0}^n 
\frac{i^{\bar n} (-i)^{n-{\bar n}} }{{\bar m}!(m-{\bar m})!{\bar n}!(n-{\bar n})!\,\hbar ^n}\, \\ \times  
\Qbracket{\Bigg}{ 
T_- 
\Sbracket{\bigg}{
\int dt \eta\FP (t)\Heis{Q}(t) 
}^{\bar m}
\Sbracket{\bigg}{
\int dt j(t)\Heis{Q}(t) 
}^{\bar n}
\, \\ \times  T_+ 
\Sbracket{\bigg}{
\int dt \eta\FM (t)\Heis{Q}(t) 
}^{m - \bar m}
\Sbracket{\bigg}{
\int dt j(t)\Heis{Q}(t) 
}^{n-\bar n}
} . 
}{
% \nonumber % 
\eqlabel{DmnGenSer0} 
}% 
%+++++++++++++++++++++++++++++++++++++++++++++
By making use of the shorthand notation (\ref{eq:Shorthand}) we can rewrite this as 
%%\begin{widetext} 
\begin{multline} % \preprintmargin
\Phi \R(\eta;j) =  
%1\, + \\ 
1 + 
\sum_{m+n\geq 1} \sum_{{\bar m}=0}^m \sum_{{\bar n}=0}^n 
\frac{i^{\bar n} (-i)^{n-{\bar n}} }{{\bar m}!(m-{\bar m})!{\bar n}!(n-{\bar n})!\,\hbar ^n} 
\, \\ \times 
\left(\int dt\right)^{m+n} 
\eta (t_1)\cdots\eta (t_m)
 j  (t'_1) \cdots  j  (t'_n)
%\, \\ \times 
\Qbracket{\Big}{ 
T_-
\Heis{Q}^{(-)}(t_1)\cdots\Heis{Q}^{(-)}(t_{\bar m})
\, \\ \times 
\Heis{Q}(t'_1)\cdots\Heis{Q}(t'_{\bar n})
\, T_+
\Heis{Q}^{(+)}(t_{{\bar m}+1})\cdots\Heis{Q}^{(+)}(t_m)
\Heis{Q}(t'_{{\bar n}+1})\cdots\Heis{Q}(t'_n)}
, 
\eqlabel{DmnGenSer} % \preprintmargin
\end{multline}% 
where all time integrations are now explicit. 
By using the definition of the response function, 
cf.\ \Eq{DRDef},  
we then find 
%%\begin{widetext} 
\begin{multline} % \preprintmargin
{\mathcal D}\R^{(m;n)}(t_1,\cdots,t_m;t'_1,\cdots,t'_n ) = 
\left(
 -\frac{i}{\hbar }
\right)^n {\sum_{\text{perm}}} \sum_{\bar m=0}^m \sum_{\bar n=0}^n  \frac{(-1)^{{\bar n}} }{{\bar m}!(m-{\bar m})!{\bar n}!(n-{\bar n})!} 
\\ \times  \Qbracket{\Big}{ 
T_-
\Heis{Q}^{(-)}(\tau_1)\cdots\Heis{Q}^{(-)}(\tau_{\bar m})
\Heis{Q}(\tau'_1)\cdots\Heis{Q}(\tau'  _{\bar n})
\, \\ \times 
T_+\Heis{Q}^{(+)}(\tau _{\bar m+1})\cdots\Heis{Q}^{(+)}(\tau_m)
\Heis{Q}(\tau'_{\bar n+1})\cdots\Heis{Q}(\tau'_n)
} 
. 
\eqlabel{GenDmn0} % \preprintmargin
\end{multline}% 
%%\end{widetext} 
Here, $\tau _1,\cdots,\tau _m$ and $\tau' _1,\cdots,\tau' _n$ are, 
respectively, permutations of 
$ t _1,\cdots, t _m$ and $ t' _1,\cdots, t' _n$, and 
$\Sigma _{\text{perm}} $ denotes a summation 
over a total of $m!n!$ of such permutations. 
The average on the RHS of (\ref{eq:GenDmn0}) is symmetric with respect to 
an arbitrary permutation within 
\mbox{$\tau_1,\cdots,\tau_{\bar m}$}, 
within \mbox{$\tau'_1,\cdots,\tau'  _{\bar n}$}, 
within \mbox{$\tau _{\bar m+1},\cdots,\tau_m$}, or 
within \mbox{$\tau'_{\bar n+1},\cdots,\tau'_n$}. 
Confining the summation to permutations 
resulting in different terms then cancels the factorials 
in the denominator of the prefactor. 
As a result we have  
%%\begin{widetext} 
\begin{multline} % \preprintmargin
{\mathcal D}\R^{(m;n)}(t_1,\cdots,t_m;t'_1,\cdots,t'_n ) \\ =  
\left(
 -\frac{i}{\hbar }
\right)^n  \sum_{\bar m=0}^m \sum_{\bar n=0}^n (-1)^{\bar n} 
%\\ \times  
{\sum_{\text{perm}}}' 
\Qbracket{\Big}{ 
T_-
\Heis{Q}^{(-)}(\tau_1)\cdots\Heis{Q}^{(-)}(\tau_{\bar m})
\Heis{Q}(\tau'_1)\cdots\Heis{Q}(\tau'  _{\bar n})
\, \\ \times 
T_+\Heis{Q}^{(+)}(\tau _{\bar m+1})\cdots\Heis{Q}^{(+)}(\tau_m)
\Heis{Q}(\tau'_{\bar n+1})\cdots\Heis{Q}(\tau'_n)
} 
, 
\eqlabel{GenDmn} % \preprintmargin
\end{multline}% 
%\end{widetext} 
where $\sum_{\text{perm}}'$ is a summation over all different terms of given structure, 
i.e., such that cannot be transformed one into another merely by permuting factors under the time orderings. 
For $n=0$ we find definitions of the time-normal averages identical to (\ref{eq:AlgEx}). 
The total number of terms on the RHS of (\ref{eq:GenDmn}) 
equals $2^{m+n}$ 
so that expressions for higher order response 
functions grow very bulky. 

Equation (\ref{eq:GenDmn}) does not appear to exhibit any causality properties, and, in fact, it takes some effort to prove that such properties hold. Causality conditions for ${\mathcal D}\R^{(m;n)}$ will be discussed in section \ref{ch:Caus}, where we also derive an alternative formula for the response functions making their explicitly causal nature evident. 

The overall phase factor in (\ref{eq:GenDmn}) depends only on the number of inputs; the same applies to the overall power of Planck's constant. The relative signs of terms depend on the number of ``input'' $\Heis{Q}$s under the $T_-$-ordering. This rule can be traced back to the representation of $\Phi(\eta_-,\eta_+)$ as a Schwinger amplitude, cf.\ \Eq{SSAv}.
It is obvious that if we neglect noncommutativity of the 
operators, the sign factor makes all 
${\mathcal D}^{(m;n)}$ with $n>1$, that is, 
the response functions proper, zeros. 
This is a direct consequence of equation (\ref{eq:FjEqF}). 
It clearly shows that commutators express response. 
However it would be incorrect to say that the 
response simply stops functioning in the classical limit $\hbar \to 0$. 
Nonzero powers of $\hbar $ are  
found in the denominators 
of the quantum formulae for the response 
functions, cf.\ \Eq{GenDmn}; 
in the classical limit, these formulae 
do not give zeros but indeterminates $\displaystyle \frac{0}{0} $. 
Correct classical limit can only be achieved by analysing physical details and not by formal means like neglecting noncommutativity of operators or setting Planck's constant to zero.

We illustrate equation (\ref{eq:GenDmn}) by constructing a formula 
for ${\mathcal D}\R^{(2,1)}$. 
From (\ref{eq:DRDef}), 
\begin{align} 
\begin{aligned}
{\mathcal D}\R^{(2;1)}(t_1,t_2;t') = 
\frac{\delta \qav{{\mathcal T}\bm{:}\Heis{Q}_j(t_1)\Heis{Q}_j(t_2)\bm{:}}}{\delta j(t')}
\bigg |_{j=0} . 
\end{aligned}%
% \nonumber % 
\eqlabel{DR21} 
\end{align}%
This is the simplest example of a 
quantum-statitistical  
response function in the true sense of the term. 
It does not follow from Glauber's photodetection theory nor 
from Kubo's 
linear response theory. 
For $D\R^{(2;1)}$ equation (\ref{eq:GenDmn}) gives  
%\begin{widetext} 
%=============================================
\eqA{ 
{\mathcal D}\R^{(2,1)}(t_1,t_2;t') \A = -\frac{i}{\hbar } 
\int dt_1' dt_2' \bigg\{
\delta^{(+)}(t_1-t_1')\delta^{(+)}(t_2-t_2')
\\ \A \ \ \ \ \ \times % \hfill 
\Sbracket{\Big}{
\qav{T_+\Heis{Q}(t'_1)\Heis{Q}(t'_2){\Heis{Q}}(t')} - 
\qav{{\Heis{Q}}(t') T_+\Heis{Q}(t'_1)\Heis{Q}(t'_2)} 
} 
\\ \A \ \ \ + \, %\hspace{0.075\columnwidth}\hspace{0.05\columnwidth}
\delta^{(-)}(t_1-t_1')\delta^{(+)}(t_2-t_2')
\\ \A \ \ \ \ \ \times % \hfill  
\Sbracket{\Big}{
\qav{\Heis{Q}(t'_1) T_+\Heis{Q}(t'_2){\Heis{Q}}(t')} -  
\qav{T_-\big[\Heis{Q}(t'_1){\Heis{Q}}(t')\big] \Heis{Q}(t'_2)} 
} 
%\\ 
%%
%\delta^{(+)}(t_1-t_1')\delta^{(-)}(t_2-t_2')
%\Sbracket{\Big}{
%\qav{\Heis{Q}(t'_2) T_+\Heis{Q}(t'_1){\Heis{Q}}(t')} - 
%\qav{T_-\big[\Heis{Q}(t'_2){\Heis{Q}}(t')\big] \Heis{Q}(t'_1)} 
%}\,   + \\ 
%% 
%\delta^{(-)}(t_1-t_1')\delta^{(-)}(t_2-t_2')
%\Sbracket{\Big}{
%\qav{T_-\big[\Heis{Q}(t'_1)\Heis{Q}(t'_2)\big]{\Heis{Q}}(t')} - 
%\qav{T_-\Heis{Q}(t'_1)\Heis{Q}(t'_2){\Heis{Q}}(t')} 
%} 
\bigg\} 
\\ \A \ \ \ + \textrm{c.c.}\ . 
}{
% \nonumber % 
\eqlabel{D21} 
}% 
In this expression we retain the orderings only where applicable; in the last term we put square brackets  around $\Heis{Q}(t'_1){\Heis{Q}}(t')$ to delineate the range to which the $T_-$-ordering is applied. To emphasise the actual structure of (\ref{eq:GenDmn}), in equation (\ref{eq:D21}) we did not use the shorthand (\ref{eq:Shorthand}). 
%*******************************************
\subsection{Expressing Green's functions by the response functions}
%*******************************************
Description of the system in terms of the response 
functions is by construction equivalent to the conventional 
QFT description in terms of Green's functions. 
By using the inverse substitution (\ref{eq:SubstEta}) we can write 
\begin{gather} 
\begin{aligned}
\Phi (\eta _-,\eta _+) & = \Phi\R (\eta ,\sigma )|_{
\eta = i\eta _- - i\eta _+, 
\sigma = \hbar \eta _+^{(+)}+\hbar \eta _-^{(-)}
} . 
\end{aligned}%
% \nonumber % 
\eqlabel{BackToTpm} 
\end{gather}%
This way, not only the response functions may be expressed by Green's 
functions using (\ref{eq:ToPhiR}), but also Green's 
functions may be expressed by the response functions 
using (\ref{eq:BackToTpm}). 
Proceeding in close similarity to how \Eq{GenDmn} was derived we find  
%%\begin{widetext} 
\begin{multline} % \preprintmargin
\qav{
T_- \Heis{Q}(t_1)\cdots\Heis{Q}(t_k) 
\,
T_+ \Heis{Q}(t'_1)\cdots\Heis{Q}(t'_l) 
} =  \sum_{\bar k = 0}^k\sum_{\bar l = 0}^l 
(-i)^{k-\bar k} i^{l-\bar l} 
\hbar ^{k+l-\bar k - \bar l}
\\ \times  
{\sum_{\text{perm}}}' 
{\mathcal D}^{(\bar k + \bar l;k+l-\bar k - \bar l)}\R(
\tau _1,\cdots,\tau _{\bar k},
\tau '_1,\cdots,\tau '_{\bar l} ; 
\tau ^{(+)}_{\bar k+1},\cdots,\tau ^{(+)} _{k},
\tau ^{\prime(-)}_{\bar l+1}\cdots,\tau ^{\prime(-)}_{l} ) .   
\eqlabel{TpmByDmn} % \preprintmargin
\end{multline}% 
%\end{widetext}%
Following the pattern of shorthand (\ref{eq:Shorthand}), we 
introduced here another shorthand  
%=============================================
\eqA{ 
{\mathcal D}\R^{(m;n)}(\cdots t^{(\pm)}\cdots ) = %\\ 
\int d t' \delta^{(\pm)}(t - t') 
{\mathcal D}\R^{(m;n)}(\cdots t'\cdots ). 
}{
% \nonumber % \eqlabel{} 
}% 
%+++++++++++++++++++++++++++++++++++++++++++++
Times and summations in \Eq{TpmByDmn}) have the same meaning as in (\ref{eq:GenDmn}). 

It is instructive to isolate in \Eq{TpmByDmn} the term with $\bar k=k$, 
$\bar l =l$. This terms equals 
%=============================================
\eqA{ 
{\mathcal D}\R^{(k+l;0)}(t_1,\cdots,t_k,t'_1,\cdots,t'_l) = %\\ 
\qav{{\mathcal T}\bm{:}
\Heis{Q}(t_1)\cdots\Heis{Q}(t_k)
\Heis{Q}(t'_1)\cdots\Heis{Q}(t'_l)
\bm{:}}.   
}{
% \nonumber % \eqlabel{} 
}% 
%+++++++++++++++++++++++++++++++++++++++++++++
This is the only term on the RHS of (\ref{eq:TpmByDmn}) 
that enters with zero power of Planck's constant. 
Other terms are expressed by response functions in the true sense 
of the word which all enter with nonzero powers of Planck's constant.
This can be written as a symbolic relation  
%%\begin{widetext} 
\begin{multline} % 
\preprintmargin
\qav{
T_- \Heis{Q}(t_1)\cdots\Heis{Q}(t_k) 
\,
T_+ \Heis{Q}(t'_1)\cdots\Heis{Q}(t'_l) 
} \\ - 
\qav{{\mathcal T}\bm{:}
\Heis{Q}(t_1)\cdots\Heis{Q}(t_k)
\Heis{Q}(t'_1)\cdots\Heis{Q}(t'_l)
\bm{:}}  = %\\ 
\hbar \times \text{{\em \{Response\}\/}}. 
\eqlabel{TpmByDmn1} % 
\preprintmargin
\end{multline}% 
%%\end{widetext}%
The difference on the LHS of \Eq{TpmByDmn1}   
may be nonzero only because the operators do not commute. 
This equation is  
another manifestation of the general link between noncommutativity  
and response. 

The terms in the sum denoted symbolically as {\em \{Response\}\/} may contain 
additional powers of $\hbar $; 
some of them may have classical 
interpretation and others may not. 
Importantly, the hierarchy of powers of $\hbar $ here is not the hierarchy 
of quantum corrections. 
Indeed, the power of Planck's constant
scales simply as the the order of nonlinearity. 
The maximal power of $\hbar $ have the terms with $k+l-1$ inputs and 
one output. 
All these terms are expressed by the same nonlinear 
{\em non-statistical\/} response function 
%=============================================
\eqA{ 
{\mathcal D}\R^{(1;k+l-1)}(t;t'_1,\cdots,t'_{k+l-1}) = %\\ 
\frac{\delta^{k+l-1}\qav{\Heis{Q}_j(t)}}
{\delta j(t'_1)\cdots\delta j(t'_{k+l-1})} 
\, \bigg|_{j=0} ,
}{
\eqlabel{DRnl} % 
% \nonumber % \eqlabel{} 
}% 
%+++++++++++++++++++++++++++++++++++++++++++++
where we have used that ${{\mathcal T}\bm{:}
\Heis{Q}(t)
\bm{:}}= \Heis{Q}(t)$, cf.\ \Eq{AlgEx}. The function (\ref{eq:DRnl}) may well have a purely classical meaning (for instance, 
in the mean-field approximation). 
This is just another example of how dangerous it is to formally classify 
quantum effects by the powers of Planck's constant. 

We complete this paragraph with some examples. 
For $k+l=1$ \Eq{TpmByDmn} reduces to 
\begin{gather} 
\begin{aligned}
\qav{T_- \Heis{Q}(t)} = 
\qav{T_+ \Heis{Q}(t)} = 
\qav{\Heis{Q}(t)}
.   
\end{aligned}%
% \nonumber % \eqlabel{} 
\end{gather}% 
For $k=0$, $l=2$: 
\begin{multline} % 
\preprintmargin
\qav{
T_+ \Heis{Q}(t_1)\Heis{Q}(t_2)} - 
\qav{{\mathcal T}\bm{:}\Heis{Q}(t_1)\Heis{Q}(t_2)
\bm{:}} \\ =  i\hbar \left[
{\mathcal D}\R^{(1;1)}(t_1;t^{(-)}_2) + {\mathcal D}\R^{(1;1)}(t_2;t^{(-)}_1)
\right]  . 
\eqlabel{T02} % 
\preprintmargin
\end{multline}% 
This is a nonlinear counterpart of \eqOscDFbyDR. 
This becomes evident if the system is homogeneous in time. 
In this case 
\begin{gather} 
\begin{aligned}
{\mathcal D}\R^{(1;1)}(t_1;t_2^{(-)}) = {\mathcal D}\R^{(1;1)(+)}(t_1-t_2),  
\end{aligned}%
% \nonumber % \eqlabel{} 
\end{gather}%
and similarly for the second term in (\ref{eq:T02}). 
For $k=l=1$, and assuming time homogeneity we find 
\begin{multline} % 
\preprintmargin
\qav{
\Heis{Q}(t_1)\Heis{Q}(t_2)} - 
\qav{{\mathcal T}\bm{:}\Heis{Q}(t_1)\Heis{Q}(t_2)
\bm{:}} \\ =  i\hbar \left[
{\mathcal D}\R^{(1;1)(+)}(t_1-t_2) - {\mathcal D}\R^{(1;1)(-)}(t_2-t_1)
\right]  ,  
\eqlabel{T11} % 
\preprintmargin
\end{multline}% 
which is a nonlinear counterpart of \eqOscDbyDR. 
An interesting observations is that the RHS and hence the LHS of 
(\ref{eq:T11}) remain frequency-positive with respect to 
$t_1-t_2$ also in the general case of an interacting system. 
%******************************************* 
\section{Causality of the quantum response}\chlabel{ch:Caus} 
%******************************************* 
\subsection{Causality conditions for the response functions}%
\chlabel{ch:CausCond} 
%******************************************* 
It is important to realise that nothing in the above warrants 
causality in the response formulation. Formula (\ref{eq:GenDmn}) involves separation of the frequency-positive 
and negative parts of Green's functions (\ref{eq:FGNL}). 
In turn this implies smearing of the 
latter from $t=-\infty$ to $t=+\infty$. 
Therefore all terms in (\ref{eq:GenDmn}) 
contain a contribution from the future. 
This ``future tail'' could well make the ${\mathcal D}\R^{(m;n)}$s 
acausal, or only approximately causal. 
We now show that in fact  
the ``future tail'' cancels {\em exactly\/}. 

To understand what kind of causality condition is to be expected, 
we employ the structural analogy between the quantum and classical response formulations discussed in section \ref{ch:PhysPr}. 
The classical response functions (\ref{eq:DCDef}) must vanish if 
at least one input time exceeds all output times. 
By analogy with the response of a classical stochastic 
system we formulate 
the causality condition for ${{\mathcal D}}\R ^{(m,n)}$ as,  
%%\begin{widetext} 
%=============================================
\eqA{ 
{{\mathcal D}}^{(m,n)}_{\text{R}} 
(t_1,\cdots,t_m;t'_1,\cdots,t'_n) 
\A = 0, \A  
\max(t_1,\cdots,t_m)\A <\max(t'_1,\cdots,t'_n) 
. 
}{
\eqlabel{CausDmn} 
}% 
%+++++++++++++++++++++++++++++++++++++++++++++
By the same analogy we expect that
%=============================================
\eqA{ 
{{\mathcal D}}^{(0,n)}_{\text{R}}(;t'_1,\cdots,t'_n) \A= 0,\A n\A>0.  
}{
\eqlabel{CausD0n} 
}% 
%+++++++++++++++++++++++++++++++++++++++++++++
This is a quantum analogue of the classical condition 
%=============================================
\eqA{ 
{{\mathcal D}}^{(0,n)}_{\text{C}}(;t'_1,\cdots,t'_n) \A=    
\frac{\delta^n\av{1}
}{\delta j(t'_1)\cdots\delta j(t'_n)}  |_{j=0}   
 = 0,\ \ \   n>0, 
}{
\eqlabel{CausDC0n} 
}% 
%+++++++++++++++++++++++++++++++++++++++++++++
required by conservation of probability. 
Indeed, $\av{1} = 1$ is a statement that full probability is one;  
(\ref{eq:CausDC0n}) stipulates that full probability does not 
depend on the external current, in other words, is conserved. 
In fact, equation (\ref{eq:CausD0n}) 
follows directly from (\ref{eq:DRDef}) 
by noting that $\Phi \R(0 ;j)=\qav{1} = \Tr \rho _0 = 1$ which is full probability in quantum mechanics. 
Equation (\ref{eq:CausD0n}) thus expresses conservation of probability 
in quantum mechanics.

We note that a general causality condition for the response functions 
may only be nonrelativistic. 
Claiming otherwise would mean, for instance, that 
an approach based on the 
nonrelativistic Schr\"odinger equation can be made  
relativistically causal merely by rewriting it in response terms. 
This simple example emphasises that 
relativistic causality is prone 
to nonrelativistic approximations in dynamics and can therefore 
only be demonstrated within a truly relativistic model. 
%******************************************* 
\subsection{Causality of Kubo's linear response}\chlabel{ch:CausKubo} 
%******************************************* 
It is instructive to trace how causality emerges in the linear response function. By employing the general formula (\ref{eq:GenDmn}) we have 
%=============================================
\eqM{ 
{\mathcal D}\R^{(1;1)}(t;t') = -\frac{i}{\hbar} \int d\bar t \\ \times  
\Big[ 
\delta^{(+)}(t-\bar t)\qav{T_+ \Heis{Q}(\bar t)\Heis{Q}(t')}
- \delta^{(+)}(t-\bar t)\qav{\Heis{Q}(t')\Heis{Q}(\bar t)}
\\ + \delta^{(-)}(t-\bar t)\qav{\Heis{Q}(\bar t)\Heis{Q}(t')}
- \delta^{(-)}(t-\bar t)\qav{T_-\Heis{Q}(t')\Heis{Q}(\bar t)}
\Big ] . 
}{
% \nonumber % \eqlabel{} 
}% 
%+++++++++++++++++++++++++++++++++++++++++++++
If we divide the whole integration region into $\bar t>t'$ 
and $\bar t<t'$, only the former contributes and we find 
%=============================================
\eqM{ 
{\mathcal D}\R^{(1;1)}(t;t') = -\frac{i}{\hbar} \int d\bar t \, 
\theta (\bar t-t')%\,\\ \times  
\bigg\{
\Big[\delta^{(+)}(t-\bar t) + \delta^{(-)}(t-\bar t)\Big]
\qav{\Heis{Q}(\bar t)\Heis{Q}(t')}
\\ - \Big[\delta^{(+)}(t-\bar t) + \delta^{(-)}(t-\bar t)\Big]
\qav{\Heis{Q}(t')\Heis{Q}(\bar t)}
\bigg\}
. 
}{
\eqlabel{KuboRaw} % 
}% 
%+++++++++++++++++++++++++++++++++++++++++++++
By noting that 
%=============================================
\eqA{ 
\delta^{(+)}(t-\bar t)+ \delta^{(-)}(t-\bar t) = 
\delta(t-\bar t), 
}{
% \nonumber % \eqlabel{} 
}% 
%+++++++++++++++++++++++++++++++++++++++++++++
the integration over $\bar t$ may be performed explicitly. Kubo's linear response function (\ref{eq:FieldKuboA}) clearly follows. 

This example emphasises the key point: the terms in \Eq{GenDmn} 
can be pairwise combined so that the latest integration time becomes associated with an ``intact'' $\delta$-function making the overall order of times meaningful. 
This idea will be used in the general proof of causality given in the next paragraph.
%******************************************* 
\subsection{Direct proof of the causality conditions% 
\chlabel{ch:CausProof}}
%******************************************* 
Equations (\ref{eq:CausDmn}) and (\ref{eq:CausD0n}) may be traced down to the following simple structural property of the double time ordering. 
Let $\Pi $ denote an arbitrary product of the field operators 
where all time arguments are less than some $t$. Then,  obviously, 
\begin{align} 
\begin{aligned}
T_-[\Pi  \Heis{Q}(t)] & = T_-[\Pi ]\Heis{Q}(t), 
\\ 
T_+[\Heis{Q}(t) \Pi] & = \Heis{Q}(t)T_+[\Pi  ] 
. 
\end{aligned}%
% \nonumber % 
\eqlabel{LatestTInTpTm} 
\end{align}%
Square brackets here specify the range to which the time orderings are applied. 
The latest field operator in 
a double-time-ordered product can therefore 
equally be placed under the $T_+$ 
or $T_-$ ordering: 
%=============================================
\eqA{ 
T_-[\Pi _- \Heis{Q}(t)]\, T_+ [\Pi _+]  
= T_-[\Pi _- ]\, T_+ [\Heis{Q}(t)\Pi _+] 
= T_-[\Pi _- ]\Heis{Q}(t)\, T_+ [\Pi _+]  
,  
}{
\eqlabel{LatestTInTPM} 
% \nonumber % \eqlabel{} 
}% 
%+++++++++++++++++++++++++++++++++++++++++++++
where $\Pi _+ $ and $\Pi _-$ are now 
two independent operator products 
with all times preceding $t$.  
In turn, this means that in the functional Taylor expansion of 
$\Phi (\eta _-,\eta _+)$ 
with respect to $\eta _+,\eta _-$ the latest operator in a 
double-time-ordered operator average always enters in the combination: 
\begin{multline} % 
\preprintmargin
- i 
\big[ 
\eta _+(t) - \eta _-(t)
\big] 
\qav{T_-[\Pi _- ]\Heis{Q}(t)\, T_+ [\Pi _+]} 
\\ =   
\eta (t)
\qav{T_-[\Pi _- ]\Heis{Q}(t)\, T_+ [\Pi _+]} .  
\eqlabel{FieldCausComb} 
\preprintmargin
\end{multline}% 
so that the latest time is always associated with an ``intact'' $\eta $ 
without any admixture of $\sigma $. ``Intact'' means that the latest 
$\eta $ is not split into the frequency-positive and negative parts, 
making the time order of arguments meaningful. 

To turn these leading considerations into a proof we have to show 
that\linebreak
$- i 
\eta _+(t)
\qav{T_-[\Pi _- ]\Heis{Q}(t)\, T_+ [\Pi _+]} 
$ and 
$i 
\eta _-(t)
\qav{T_-[\Pi _- ]\Heis{Q}(t)\, T_+ [\Pi _+]} 
$ enter with\linebreak equal weight. 
We do this by manipulating the range of time integrations in 
(\ref{eq:PhiDef}). 
By making use of the symmetry of the integrand  
we may confine the time integrations to 
\mbox{$t_1<\cdots<t_m$}, \mbox{$t'_1<\cdots<t'_n $}. 
This cancels the factorials in the denominator 
of the prefactor. 
We then further split the integration range into  
$t_m>t'_n$ and $t_m<t'_n$. If $t_m>t'_n$, we define $t = t_m$, $ k = m-1$
and $ l =n$. The integration region is then
\begin{align} 
\begin{gathered}
-\infty<t_1<\cdots<t_{ k}<t<\infty,\\  
-\infty<t'_1<\cdots<t'_{ l}<t<\infty , 
\end{gathered}%
% \nonumber % 
\eqlabel{OrdTimeInt} 
\end{align}%
while the prefactor becomes $i^{ k + 1} (-i)^{ l}$. 
If $t_m<t'_n$, we define $t = t'_n$, $ k = m$
and $ l =n-1$. The integration region is then again 
defined by (\ref{eq:OrdTimeInt}), while the prefactor equals $i^{ k} (-i)^{ l + 1}$.
Furthermore, due to the ordering of the integration times, 
the $T_{\pm}$ orderings may be performed explicitly. 
The double-time-ordered products for both cases look the same:
\begin{align} 
\begin{aligned}
\qav{\Heis{Q}(t_1)\cdots\Heis{Q}(t_{ k})\Heis{Q}(t)
\Heis{Q}(t'_{ l})\cdots\Heis{Q}(t'_1)} ,  
\end{aligned}%
% \nonumber % \eqlabel{} 
\end{align}%
cf.\ \Eq{LatestTInTPM}. 
For every $ k,  l$ pair, we have two terms with $\Heis{Q}(t)$ 
coming, respectively, from the $T_+$ and $T_-$-ordered products. 
Combining them we get 
%%\begin{widetext} 
\begin{multline} % 
\preprintmargin
\Phi (\eta _-,\eta _+) = 1 - %\\ 
i  \sum_{ k+ l\geq 0} 
i^{ k} (-i)^{ l} 
\left(\int dt\right)^{ k+ l + 1} \\ \times 
\big[\eta _+(t) - \eta _-(t)\big] %\\ \times 
\eta _-(t_1)\cdots\eta _-(t_{ k})
\eta _+(t'_1)\cdots\eta _+(t'_{ l}) \\ \times  
\qav{\Heis{Q}(t_1)\cdots\Heis{Q}(t_{ k})\Heis{Q}(t)
\Heis{Q}(t'_{ l})\cdots\Heis{Q}(t'_1)}
, 
\eqlabel{PhiCaus} % 
\preprintmargin
\end{multline}% 
%%\end{widetext}%
where the time integrations are confined to (\ref{eq:OrdTimeInt}). 
It is convenient to restore the symmetry of time integrations among $t_1,\cdots,t_k$ and among $t'_1,\cdots,t_l$, resulting in 
%\begin{widetext} 
%=============================================
\eqMW{ 
\Phi (\eta _-,\eta _+) = 1 - %\\ 
i  \sum_{ k+ l\geq 0} 
\frac{i^{ k} (-i)^{ l}}{k!l!}  
\int dt %\\ \times 
\big[\eta _+(t) - \eta _-(t)\big] \\ \times  
\Qbracket{\Bigg}{ 
T_-\Sbracket{\bigg}{
\int_{-\infty}^t dt' \eta _-(t')\Heis{Q}(t')
}^k \, \Heis{Q}(t) \,  %\\ \times  
T_+ \Sbracket{\bigg}{
\int_{-\infty}^t dt' \eta _+(t')\Heis{Q}(t')
}^l 
} 
.  
}{
% \nonumber % \eqlabel{} 
}% 
%+++++++++++++++++++++++++++++++++++++++++++++
On applying substitution (\ref{eq:SubstSigma}) to this formula we find  
%=============================================
\eqMW{ 
\Phi (\eta _-,\eta _+) = 1 + 
\sum_{ k+ l\geq 0} 
\frac{1}{k!l!}  
\int dt \,\eta (t) \\ \times  
\Qbracket{\Bigg}{ 
T_-\Cbracket{\bigg}{
\int_{-\infty}^t dt' 
\Sbracket{\bigg}{
\eta^{(+)}(t') + \frac{i\sigma (t')}{\hbar } 
}
\Heis{Q}(t')
}^k \, \Heis{Q}(t) \,  \\ \times  
T_+ \Cbracket{\bigg}{
\int_{-\infty}^t dt' \Sbracket{\bigg}{
\eta^{(-)}(t') - \frac{i\sigma (t')}{\hbar } 
}
\Heis{Q}(t')
}^l 
} 
.  
}{
% \nonumber % 
\eqlabel{PhiCausF} 
}% 
%+++++++++++++++++++++++++++++++++++++++++++++
Except for one additional time integration and the integration limits, this formula is identical to \Eq{GenDmnXX} and may be manipulated in a similar way, resulting in an explicitly causal formula for the response functions, 
%=============================================
\eqMW{ 
{\mathcal D}\R^{(m+1;n)}(t_1,\cdots,t_{m+1};t'_1,\cdots,t'_n ) \\ =  
\left(
 -\frac{i}{\hbar }
\right)^n  \sum_{\bar m=0}^m \sum_{\bar n=0}^n (-1)^{\bar n} 
\,\theta(\tau -t'_1) \cdots \theta(\tau -t'_n)
\, \\ \times  
{\sum_{\text{perm}}}' 
\Qbracket{\Big}{ 
T_-
\Heis{Q}^{(-,\tau )}(\tau_1)\cdots\Heis{Q}^{(-,\tau )}(\tau_{\bar m})
\Heis{Q}(\tau'_1)\cdots\Heis{Q}(\tau'  _{\bar n})
\, \Heis{Q}(\tau ) \, 
\, \\ \times  
T_+\Heis{Q}^{(+,\tau )}(\tau _{\bar m+1})\cdots\Heis{Q}^{(+,\tau )}(\tau_m)
\Heis{Q}(\tau'_{\bar n+1})\cdots\Heis{Q}(\tau'_n)
} 
,  
}{
% \nonumber % 
\eqlabel{GenDmnCaus} 
}
%+++++++++++++++++++++++++++++++++++++++++++++
where 
we have introduced a generalisation of shorthand (\ref{eq:Shorthand})
\begin{gather} 
T_{\pm} \big[\cdots\Heis{Q}^{(\pm,t)}(t')\cdots\big]  
=  %\\ 
\int_{-\infty}^t dt'' \delta^{(\pm)} (t'-t'') 
T_ {\pm}\big[\cdots\Heis{Q}(t'')\cdots\big]   
.  
% \nonumber % 
\eqlabel{ShorthandX} 
\end{gather}%
%%\end{widetext}% 
Other notation in (\ref{eq:GenDmnCaus}) is as follows: $\tau ,\tau _1,\cdots,\tau _m$ is a permutation of $t_1,\cdots,t_{m+1}$, $\tau '_1,\cdots,\tau '_n$ is a permutation of $t'_1,\cdots,t'_n$, and the summation is over all different terms of given structure, i.e., such that cannot be transformed one into another merely by permuting factors under the time orderings. 

Equation (\ref{eq:GenDmnCaus}) is a sum of terms such that in each term one output argument is bound to be larger than all input arguments, so that causality condition (\ref{eq:CausDmn}) clearly follows. 
Due to the ``time smearing'' occuring in (\ref{eq:ShorthandX}),  this selected output argument is not bound to be larger than other output arguments, but this is irrelevant. We also see that expansion (\ref{eq:PhiCausF}) lacks terms without $\eta$s. This proves \Eq{CausD0n}.

We conclude this paragraph by constructing an explicitly causal formula 
for ${\mathcal D}\R^{(2,1)}$ alternative to (\ref{eq:D21}). 
For $D\R^{(2;1)}$ equation (\ref{eq:GenDmnCaus}) gives 
%=============================================
\eqMW{ 
{\mathcal D}\R^{(2,1)}(t_1,t_2;t') = -\frac{i}{\hbar } \theta(t_1-t')
\int_{-\infty}^{t_1}dt_2' 
\, \\ \times  
\Cbracket{\bigg}{
\delta^{(+)}(t_2-t_2')\Sbracket{\Big}{
\qav{\Heis{Q}(t_1)T_+
\Sbracket{\big}{
{\Heis{Q}}(t')\Heis{Q}(t_2')
}} - 
\qav{{\Heis{Q}}(t')\Heis{Q}(t_1)\Heis{Q}(t_2')
} 
} 
\\ + 
\delta^{(-)}(t_2-t_2')\Sbracket{\Big}{
\qav{\Heis{Q}(t_2')\Heis{Q}(t_1){\Heis{Q}}(t')
} - 
\qav{T_-
\Sbracket{\big}{
{\Heis{Q}}(t')\Heis{Q}(t_2')
} \Heis{Q}(t_1)
} 
} 
} \\ + 
\Cbracket{\big}{
t_1 \leftrightarrow t_2
} , 
}{
% \nonumber % 
\eqlabel{D21Caus} 
}% 
%+++++++++++++++++++++++++++++++++++++++++++++ 
%\end{widetext}% 
where square brackets delineate the range to which the orderings are applied. To give more emphasis to the actual structure of (\ref{eq:GenDmnCaus}) we did not use here the shorthand (\ref{eq:ShorthandX}). 
%******************************************* 
%\newpage 
\section{Conclusion}\chlabel{ch:Conc} 
%******************************************* 
We have introduced a response formulation of an arbitrary interacting bosonic field in terms of dependence of the time-normal averages of the field operator on the external source. 
While being, by construction, equivalent to the conventional Green-function techniques of the quantum field theory, this formulation exhibits key properties of a classical response picture such as reality and explicit causality. Validity of these results outside the quantum-optical paradigm requires an amendment to the conventinal definition of the time-normal operator ordering by Glauber and Kelly and Kleiner. 
%*******************************************
\section{Acknowledgments}
%*******************************************
S.\ Stenholm wishes to thank the Institut f\"ur Quantenphysik, Universit\"at Ulm, and Prof.\ W.P.\ Schleich for generous hospitality. Financial support of the Alexander von Humboldt-Stiftung is gratefully acknowledged.
%*******************************************
 
%*******************************************

%******************************************* 
\appendix
%******************************************* 
%******************************************* 
%\newpage 
\section{The positive and negative frequencies}\chlabel{app:FPN} 
%******************************************* 
The operation of separating the frequency-positive and negative 
parts of a function was defined by \Eq{FPNDef}. 
Here we discuss some properties of this operation. 
Obviously, 
\begin{align} 
\begin{aligned}
g^{(+)}(t)+g^{(-)}(t) = g(t). 
\end{aligned}%
% \nonumber % 
\eqlabel{FPNSum} 
\end{align}%
This relation, however, implies that $g_{\omega }$ is smooth at $\omega = 0$. 
An important property of (\ref{eq:FPNDef}) is the connection between 
the frequency-positive and negative 
parts and the convolution: 
\begin{multline} % 
\preprintmargin
\int dt' h^{(\pm)}(t-t') g(t') = 
\int dt' h(t-t') g^{(\pm)}(t') \\ =  
\int dt' h^{(\pm)}(t-t') g^{(\pm)}(t')  
.  
\eqlabel{FPNConv} 
\preprintmargin
\end{multline}% 
It is proved by noting that the Fourier-transform of any 
of the three expressions 
is $\theta(\pm \omega )g_{\omega }h_{\omega }$. 
With $h(t) = \delta(t)$ this conveniently expresses separation 
of the frequency-positive and negative 
parts of a function as an integral operation,  
%=============================================
\eqA{ 
g^{(\pm)}(t) = \int dt'\, \delta(t-t')g^{(\pm)}(t') = %\\ 
\int dt' \delta^{(\pm)}(t-t') g(t')  , 
}{
\eqlabel{AppFPNAsConv} 
% \nonumber % \eqlabel{} 
}% 
%+++++++++++++++++++++++++++++++++++++++++++++
where $\delta^{(\pm)}(t)$ 
are the frequency-positive and negative parts of the $\delta$-function, 
\begin{gather} 
\delta^{(\pm)}(t) = \frac{1}{2 \pi  i (\pm t - i \varepsilon) } \, . 
\eqlabel{DeltaPM} 
\end{gather}%
Obviously, 
\begin{gather} 
\delta^{(\pm)}(t) = \delta^{(\mp)}(-t), 
\eqlabel{DeltaPMInv} 
\end{gather}%
so that 
%=============================================
\eqA{ 
\int dt\, \delta^{(\pm)}(t-t')h (t) = %\\ 
\int dt\, \delta^{(\mp)}(t'-t)h (t) = 
h^{(\mp)} (t') .  
}{
\eqlabel{ShiftDl} % 
% \nonumber % \eqlabel{} 
}% 
%+++++++++++++++++++++++++++++++++++++++++++++
This formula is instrumental in deriving explicit expressions for the 
response functions; it also shows that  
%=============================================
\eqA{ 
\int dt\, h(t) g^{(\pm)}(t) 
= 
\int dtdt'\, h(t) \delta^{(\pm)}(t-t') g(t') 
= %\\ 
\int dt\, h^{(\mp)}(t) g(t) 
. 
}{
\eqlabel{ShiftPN} % 
% \nonumber % \eqlabel{} 
}% 
%+++++++++++++++++++++++++++++++++++++++++++++

Notation (\ref{eq:FPNDef}) becomes ambiguous for  
complicated expressions and composite functions like $g(h(t))$. 
Taking the frequency-positive and negative parts of an arbitrary expression with respect to $t$ will therefore be denoted as $[\cdots]^{(\pm)}_t$. 
This way, 
\begin{gather} 
\begin{aligned}
{[g(h(t))]}^{(\pm)}_t =  
\int dt' \delta^{(\pm)}(t-t') g(h(t')) ,  
\end{aligned}%
% \nonumber % \eqlabel{} 
\end{gather}%
as opposed to  
\begin{gather} 
g^{(\pm)}(h(t)) = 
\int dt' \delta^{(\pm)}(h(t)-t') g(t')  .    
% \nonumber % \eqlabel{} 
\end{gather}%
In general, taking the frequency-positive and negative parts 
of something implies $[\cdots]^{(\pm)}_t$.  
This is consistent with (\ref{eq:FPNDef}) because, obviously, 
\begin{gather} 
\begin{aligned}
{[g(t)]}^{(\pm)}_t = g^{(\pm)}(t). 
\end{aligned}%
% \nonumber % \eqlabel{} 
\end{gather}%

Both time inversion and complex conjugation replace $\e{-i\omega t}
\to \e{i\omega t}$; hence $g^{(+)}(-t)$ and $g^{(+)*}(t)$ are 
frequency-negative while $g^{(-)}(-t)$ and $g^{(-)*}(t)$ are 
frequency-positive. 
By making use of (\ref{eq:FPNDef}) it is easy to prove that 
\begin{gather} 
{[g(-t)]}^{(\pm)}_t = g^{(\mp)}(-t), \ \ 
{[g^*(t)]}^{(\pm)}_t = g^{(\mp)*}(t) 
. 
\eqlabel{FPNInv} 
\end{gather}%
These relations hold for arbitrary $g(t)$. For an even $g(t) = g(-t)$ 
we then have, 
\begin{align} 
\begin{aligned}
g^{(\pm)}(-t) = g^{(\mp)}(t), 
\end{aligned}%
% \nonumber % 
\eqlabel{FPNEven} 
\end{align}%
while for a real $g(t) = g^*(t)$, 
\begin{align} 
\begin{aligned}
g^{(\pm)*}(t) = g^{(\mp)}(t).  
\end{aligned}%
% \nonumber % 
\eqlabel{FPNReal} 
\end{align}%
%******************************************* 
%\newpage 
\section{Response of charged bosons}\chlabel{app:ChBos} 
%******************************************* 
Extention of the results to charged bosonic fields reduces in essence 
to doubling all arguments in the relations derived for 
neutral bosons by applying the shorthands (\ref{eq:HintRedef}) 
and (\ref{eq:EtaQRedef}). 
The ``charged'' analog of $\Phi (\eta _-,\eta _+)$ is given 
by (\ref{eq:DefPhiC}). A similar formula with $\Heis{F}_j,\Heis{F}_j\dg$ 
in place of $\Heis{F},\Heis{F}\dg$ defines the analog of 
$\Phi (\eta _-,\eta _+;j)$, 
while the ``charged'' analog of (\ref{eq:SigPlusJ}) reads  
%=============================================
\eqA{ 
\Phi \big(\eta _-,\bar\eta _-,\eta _+,\bar\eta _+ ;j, j^* \big) = %\\ 
\Phi\left(
\eta _- + \frac{j}{\hbar } ,
\bar\eta _- + \frac{ j^* }{\hbar } ,
\eta _+ + \frac{j}{\hbar },
\bar\eta _+ + \frac{ j^* }{\hbar }
\right)   
% \nonumber % 
}{
\eqlabel{ChargedSigPlusJ} 
% \nonumber % \eqlabel{} 
}% 
%+++++++++++++++++++++++++++++++++++++++++++++
Definition of $\Phi \R(\eta ;\sigma )$ turns into 
\begin{align} 
\begin{aligned}
\Phi\R (\bar \eta ,\eta ;\sigma ,\bar \sigma ) =  
\Phi (\eta _-,\bar \eta _-,\eta _+,\bar \eta _+)|_{\cdots},  
\end{aligned}%
% \nonumber % \eqlabel{} 
\end{align}%
where the ellipsis stands for substitutions (\ref{eq:SubstSigma}) 
supplemented by 
\begin{align} 
\begin{aligned}
\bar \eta^{\mu }_+(\bm{r},t) & = 
i \bar \eta^{\mu(-)} (\bm{r},t) 
+ \frac{1}{\hbar}\bar  \sigma^{\mu }(\bm{r},t), \\    
\bar \eta^{\mu }_-(\bm{r},t) & = 
- i\bar \eta^{\mu(+)} (\bm{r},t) 
+ \frac{1}{\hbar} \bar \sigma^{\mu }(\bm{r},t) ,   
\end{aligned}%
% \nonumber % 
\eqlabel{ChargedSubstSigma} 
\end{align}%
which is exactly the set of substitutions found in \chFieldC\ for the charged noninterating bosonic field. 
Equation (\ref{eq:PhiJR0}) becomes 
\begin{align} 
\begin{aligned}
\Phi \big(\eta _-,\bar\eta _-,\eta _+,\bar\eta _+ ;j, j^* \big)|_{\cdots} & = 
\Phi\R \big(\bar \eta ,\eta ;\sigma+j ,\bar \sigma+ j^*  \big).  
\end{aligned}%
% \nonumber % 
\eqlabel{ChargedSigPlusJret} 
\end{align}%
Definitions of the time-normal averages and response functions of 
charged bosons take the form  
%\begin{widetext} 
%=============================================
\eqA{ 
\A 
{\mathcal D}\R^{(m,n;k,l)} 
\big(
t_1,\cdots,t_m;
{\bar t}_1,\cdots,{\bar t}_n;
t'_1,\cdots,t'_k;
{\bar t}'_1,\cdots,{\bar t}'_l
\big) 
\\ \A \ \ \ =  
%\hspace{0.2\columnwidth}
\left.\frac{\delta^{k+l}\qav{{\mathcal T}\bm{:} 
\Heis{F}_j(t_1) \cdots \Heis{F}_j(t_m) 
\Heis{F}_j\dg(\bar t_1) \cdots \Heis{F}_j\dg(\bar t_n) 
\bm{:}}}{
\delta j({t}'_1) 
\cdots 
\delta j({t}'_k)
\delta j^*({\bar t}'_1) 
\cdots 
\delta j^*({\bar t}'_l)} 
\right|_{j=0} \, ,
\\ \A 
\qav{{\mathcal T}\bm{:} 
\Heis{F}_j(t_1) \cdots \Heis{F}_j(t_m) 
\Heis{F}_j\dg(\bar t_1) \cdots \Heis{F}_j\dg(\bar t_n) 
\bm{:}} \\ \A \ \ \ =  
%\hspace{0.25\columnwidth} 
\left.\frac{\delta^{m+n}\Phi\R \big(\bar\eta  ,\eta ;j,j^* \big)}{
\delta\bar\eta  (t_1)
\cdots
\delta\bar\eta  (t_m)
\delta\eta ( \bar t _1)  
\cdots 
\delta\eta( \bar t _n)} 
\right|_{\eta =\bar \eta =j=j^*  =0} 
,  
}{
\eqlabel{DRBosDiff} % \preprintmargin
% \nonumber % \eqlabel{} 
}% 
%+++++++++++++++++++++++++++++++++++++++++++++
cf.\ \Eq{DRDef}. 
The formula for ${\mathcal D}\R^{(m,n;k,l)}$, derived 
in full analogy to (\ref{eq:GenDmn}), looks    
\begin{multline} % \preprintmargin
{\mathcal D}\R^{(m,n;k,l)} 
\big(
t_1,\cdots,t_m;
{\bar t}_1,\cdots,{\bar t}_n;
t'_1,\cdots,t'_k;
{\bar t}'_1,\cdots,{\bar t}'_l
\big) 
\\ =  
\left(
 -\frac{i}{\hbar }
\right)^{k+l} 
\sum_{\bar k=0}^k \sum_{\bar l=0}^l 
\sum_{\bar m=0}^m \sum_{\bar n=0}^n  
(-1)^{\bar k + \bar l} 
{{\sum_{\text{perm}}}}' 
\Big\langle 
T_-
\Heis{F}^{(-)}(\tau  _1)\cdots\Heis{F}^{(-)}(\tau  _{\bar m})
\, \\ \times  
\Heis{F}(\tau'  _1)\cdots\Heis{F}(\tau'  _{\bar k})
\Heis{F}\dgp{(-)}(\bar\tau  _1)\cdots\Heis{F}\dgp{(-)}(\bar\tau  _{\bar n})
\Heis{F}\dg(\bar\tau'  _1)\cdots\Heis{F}\dg(\bar\tau'  _{\bar l})
\\ \times  
T_+\Heis{F}^{(+)}(\tau  _{\bar m+1})\cdots\Heis{F}^{(+)}(\tau  _m)
\Heis{F}(\tau'  _{\bar k+1})\cdots\Heis{F}(\tau'  _k)
\, \\ \times  
\Heis{F}\dgp{(-)}(\bar\tau _{\bar n+ 1})\cdots\Heis{F}\dgp{(-)}(\bar\tau _n)
\Heis{F}\dg(\bar\tau' _{\bar l+1})\cdots\Heis{F}\dg(\bar\tau' _l)
\Big\rangle 
. 
\eqlabel{GenDmnCh} % \preprintmargin
\end{multline}% 
%%\end{widetext} 
Here, 
$\tau _1,\cdots,\tau_m$,  
$\bar\tau _1,\cdots,\bar\tau_n$,  
$\tau' _1,\cdots,\tau'_k$ and   
$\bar\tau' _1,\cdots,\bar\tau'_l$ are permutations of, 
respectively, 
$t_1,\cdots,t_m$, 
$\bar t_1,\cdots,\bar t_n$, 
$t'_1,\cdots,t'_k$, 
$\bar t'_1,\cdots,\bar t'_l$; 
$\Sigma '_{\text{perm}} $ is a summation 
over permutations resulting in different 
terms on the RHS of (\ref{eq:GenDmnCh}). 
Expressions for the time-normal averages of charged fields 
follow from (\ref{eq:GenDmnCh}) as a special case 
$k=l=0$. 
The response functions of charged bosons  
obey the causality properties, 
%=============================================
\eqA{ 
{\mathcal D}\R^{(m,n;k,l)}(t_1,\cdots,t_m;\bar t_1,\cdots,\bar t_n; 
t'_1,\cdots,t'_k;\bar t'_1,\cdots,\bar t'_l) = 0, \\ 
\text{max}\{t_s,\bar t_s\}< \text{max}\{t'_s,\bar t'_s\}, 
}{
\eqlabel{CausCh} 
% \nonumber % \eqlabel{} 
}% 
%+++++++++++++++++++++++++++++++++++++++++++++
%%\begin{widetext} 
\begin{gather} 
{\mathcal D}\R^{(0,0;k,l)}(;; 
t'_1,\cdots,t'_k;\bar t'_1,\cdots,\bar t'_l) = 0, 
\end{gather}%
and the reality property, 
\begin{multline} % \preprintmargin
\left[
{\mathcal D}\R^{(m,n;k,l)}(t_1,\cdots,t_m;\bar t_1,\cdots,\bar t_n; 
t'_1,\cdots,t'_k;\bar t'_1,\cdots,\bar t'_l)
\right]^*  \\ = 
{\mathcal D}\R^{(n,m;l,k)}(
\bar t_1,\cdots,\bar t_n; 
t_1,\cdots,t_m;
\bar t'_1,\cdots,\bar t'_l;
t'_1,\cdots,t'_k
)   .   
\eqlabel{ConjCh} % \preprintmargin
\end{multline}% 
%\end{widetext}%
Similar to how reality and causality properties of neutral bosons 
are those of a real c-number field, 
properties (\ref{eq:CausCh})--(\ref{eq:ConjCh}) of charged bosons 
are clearly those of a complex stochastic c-number field. 
%******************************************* 

\begin{thebibliography}{} 
%******************************************* 
\bibitem{RespOsc}
L.\ I.\ Plimak and S.\ Stenholm, {\em Annals of Physics\/}, (2008), doi:10.1016/j.aop.2007.11.013. 
\bibitem{Schleich}
Wolfgang P.\ Schleich, {\em Quantum Optics in Phase Space\/} 
(Wiley, Berlin, 2001). 
\bibitem{MandelWolf}
Leonard Mandel and Emil Wolf, {\em Optical coherence and quantum optics\/} 
(Cambridge University Press, 1995).
\bibitem{KellyKleiner}
P.L.\ Kelly and W.H.\ Kleiner, Phys.Rev.\ {\bf 136}, A316 (1964).  
\bibitem{GlauberTN}
R.J.\ Glauber, {\em Quantum Optics and Electronics\/}, 
Les Houches Summer School of Theoretical Physics 
(Gordon and Breach, New York, 1965). 
%******************************************* 
\bibitem{SchwingerC}
J.S.\ Schwinger,  
J.\ Math. Phys.\ {\bf 2}, 407 (1961); see also a review by Fred Cooper, {\em e-print\/} arXiv:hep-th/950407v1 (1995). 
\bibitem{Kubo}
R.\ Kubo, {\em Statistical Mechanics\/} (North-Holland, 1965).
\bibitem{Bogol}
N.N.\ Bogoliubov and D.V.\ Shirkov, 
{\em Introduction to the theory of quantized fields\/} 
(Wiley, New York, 1980). 
\bibitem{BWO}
L.I.\ Plimak, M.\ Fleischhauer, M.K.\ Olsen, and M.J.\ Collett, 
Phys.\ Rev.\ A {\bf 67}, 013812 (2003). 
\bibitem{Keldysh}
O.V.\ Konstantinov and V.I.\ Perel, 
Zh.\ Eksp.\ Theor.\ Phys.\ {\bf 39}, 197 (1960) 
[Sov.\ Phys.\ JETP {\bf 12}, 142 (1961)];  
L.V.\ Keldysh, {\em ibid.\/} {\bf 47}, 1515 (1964) 
[{\bf 20}, 1018 (1964)]. 
\bibitem{GlauberPhDet}
Roy J.\ Glauber, Phys.\ Rev.\ {\bf 130}, 2529 (1963). 
\bibitem{Corresp}
L.I.\ Plimak, 
Phys.\ Rev.\ A {\bf 50}, 2120 (1994). 
\bibitem{IntRange}
Whenever limits of integration are omitted, a maximal possible 
range is implied: the whole space, the whole time axis, 
and so on. 

\hide{
\bibitem{Zinn}
J.\	Zinn-Justin, {\em Quantum field theory and critical phenomena\/} 
(Oxford, Clarendon Press, 2004). 
\bibitem{Schwinger}
J.S.\ Schwinger,  
{\em Selected Papers on ``Quantum Electrodynamics''\/}
(Dover Publications, New York, 1958). 
%******************************************* 

\bibitem{Feedback}
L.I.\ Plimak, Quant.\ Semicl.\ Opt.\ {\bf 8}, 323 (1996).  
\bibitem{YamFeedB}
S.\ Machida and Y.\ Yamamoto, Opt.\ Comm.\ {\bf 57}, 290 (1986).
\bibitem{Nelson}
E.\ Nelson, Journal of Functional Analysis {\bf 12}, 97 (1973). 
\bibitem{PlimakWalls}
L.I.\ Plimak and D.F.\ Walls,
Phys.\ Rev.\ A {\bf 50}, 2627 (1994). 
\bibitem{Prepr}
L.I.\ Plimak, M.\ Fleischhauer, M.K.\ Olsen, and M.J.\ Collett, 
{\em Quantum-field-theoretical techniques for stochastic 
representation of quantum problems\/}, e-print cond-mat/0102483.
\bibitem{EuLett}
L.I.\ Plimak {\em et al\/}., Europhys.\ Lett.\ {\bf 56}, 372 (2001). 
\bibitem{AskMurray1}
M.K.\ Olsen, L.I.\ Plimak, and M.J.\ Collett, 
Phys.\ Rev.\ A {\bf 64}, 063601 (2001). 
\bibitem{AskMurray2}
M.K.\ Olsen, L.I.\ Plimak, and M.\ Fleischhauer, 
Phys.\ Rev.\ A {\bf 65}, 053806 (2002). 
\bibitem{Cumul}
R.L.\ Stratonovich, {\em Nonlinear Nonequilibrium Thermodynamics I\/} 
(Springer, Berlin, 1992).
\bibitem{Steel}
M.J.\ Steel {\em et al.\/}, 
Phys.\ Rev.\ A {\bf 58}, 4824 (1998). 
\bibitem{Hori}
T.Hori, 
Prog.\ Theor.\ Phys.\ {\bf 7}, 378 (1952). 
}
%******************************************* 
\end{thebibliography}
\end{document}